\documentclass[12pt]{article}
\usepackage[margin=1in]{geometry}
\usepackage{longtable}
\usepackage{amsmath,amssymb,booktabs,graphicx,natbib,setspace,hyperref,threeparttable,caption}
\usepackage{xcolor}
\usepackage{placeins}
\usepackage{url}

\newcommand{\todo}[1]{\textcolor{blue}{[TODO: #1]}}

\newcommand{\nCand}{988}
\newcommand{\nTierOne}{286}
\newcommand{\nCandPkg}{727}

\newcommand{\nScreened}{1,331}
\newcommand{\nScreenedCand}{988}
\newcommand{\nScreenedNonCand}{343}

\newcommand{\nScreenedText}{255}

\newcommand{\nCandExtra}{611}
\newcommand{\nScreenedExtra}{625}
\newcommand{\nScreenedCandExtra}{608}

\newcommand{\nEligibleExtra}{29}
\newcommand{\nNearMissExtra}{159}

\newcommand{\nEligible}{92}

\newcommand{\nEligibleNonCand}{2}
\newcommand{\nEligSynth}{18}

\newcommand{\nUsableNotReest}{1}

\newcommand{\nJMdeposits}{21}

\newcommand{\nPkgFiles}{1,070}
\newcommand{\nPkgNonCand}{343}

\IfFileExists{tables/results_numbers.tex}{\newcommand{\nMainSpecs}{355}
\newcommand{\nMainComparisons}{411}
\newcommand{\nMainPapers}{40}

\newcommand{\corrCSA}{0.61}
\newcommand{\corrBJS}{0.70}
\newcommand{\flipCSA}{25.1}
\newcommand{\flipBJS}{18.0}
\newcommand{\sigflipCSA}{10.2}
\newcommand{\sigflipBJS}{7.2}
\newcommand{\seratioCSA}{1.61}
\newcommand{\seratioBJS}{1.30}
\newcommand{\flipCSApw}{19.9}
\newcommand{\flipBJSpw}{16.4}

\newcommand{\corrCSApw}{0.62}
\newcommand{\corrBJSpw}{0.35}
\newcommand{\nCSA}{283}
\newcommand{\nBJS}{306}
\newcommand{\papersCSA}{28}
\newcommand{\papersBJS}{29}

\newcommand{\flipCSApost}{26.4}
\newcommand{\flipBJSpost}{16.7}

\newcommand{\nPreCSA}{44}
\newcommand{\nPreBJS}{42}

\newcommand{\preSigTWFE}{27.3}
\newcommand{\preSigCSA}{25.0}
\newcommand{\preSigBJS}{23.8}
\newcommand{\negRowsMedian}{22.4}
\newcommand{\negAbsMedian}{5.4}

\newcommand{\hCorrCSA}{0.62}
\newcommand{\hCorrBJS}{0.61}

\newcommand{\hFlipText}{28.0 percent for both estimators}
\newcommand{\hSigCSA}{13.4}
\newcommand{\hSigBJS}{11.9}
\newcommand{\shareNeverMean}{38.3}
\newcommand{\shareAlwaysMean}{4.4}
\newcommand{\medianCohorts}{14}
\newcommand{\nModPapers}{9}

\newcommand{\nTwPapers}{19}

\newcommand{\flipCSAmod}{24.6}
\newcommand{\flipCSAtw}{25.4}

\newcommand{\corrCSAmod}{0.41}
\newcommand{\corrCSAtw}{0.74}
\newcommand{\nCovCSA}{280}
\newcommand{\nCovBJS}{306}
\newcommand{\sigflipCSAzero}{9.2}
\newcommand{\sigflipBJSzero}{3.6}
\newcommand{\sigflipCSAcorr}{11.0}
\newcommand{\sigflipBJScorr}{8.2}
\newcommand{\sigflipCSAbvar}{12.0}
\newcommand{\sigflipBJSbvar}{8.8}
\newcommand{\sigdiffCSA}{25.4}
\newcommand{\sigdiffBJS}{24.8}
\newcommand{\sigdiffCSAzero}{15.2}
\newcommand{\sigdiffBJSzero}{7.2}
\newcommand{\bootCorCSA}{0.53}
\newcommand{\bootCorBJS}{0.82}
\newcommand{\nNegVarCSA}{1}
\newcommand{\nNegVarBJS}{9}
\newcommand{\seratioCSAmed}{1.36}
\newcommand{\seratioBJSmed}{1.11}

\newcommand{\corrCSAunbal}{0.47}
\newcommand{\corrBJSunbal}{0.55}
\newcommand{\corrCSAbal}{0.92}
\newcommand{\corrBJSbal}{0.99}

\newcommand{\nCSAlean}{8}

\newcommand{\nCSAunsupRowsAll}{61}
\newcommand{\nCSAunsupPapersAll}{11}
\newcommand{\nCSAunsupRows}{42}
\newcommand{\nCSAunsupPapers}{7}
\newcommand{\nCSArankdefPapers}{2}
\newcommand{\nCSAnocellsPapers}{2}

\newcommand{\nBJSnotIdent}{34}

\newcommand{\nPubCSspecs}{80}
\newcommand{\nPubCSpapers}{8}
\newcommand{\corrCSAauth}{0.60}
\newcommand{\flipCSAauth}{24.0}
\newcommand{\sigflipCSAauth}{9.2}

\newcommand{\authMedGapSE}{0.006}

\newcommand{\nAuthMoveOneSE}{5}
\newcommand{\nAuthMoveOneSEpapers}{2}

\newcommand{\corrCSAnegLow}{0.94}
\newcommand{\corrCSAnegHigh}{0.42}
\newcommand{\corrBJSnegLow}{0.99}
\newcommand{\corrBJSnegHigh}{-0.27}
\newcommand{\corrCSAnoNever}{0.33}
\newcommand{\corrBJSnoNever}{-0.05}
\newcommand{\nNoNeverPapers}{5}
\newcommand{\nNegLowCSA}{54}
\newcommand{\nNegHighCSA}{53}
\newcommand{\nNegLowBJS}{61}
\newcommand{\nNegHighBJS}{60}
\newcommand{\nNoNeverCSA}{66}
\newcommand{\nNoNeverBJS}{74}

\newcommand{\regUnbalBJS}{0.115}
\newcommand{\regUnbalBJSse}{0.060}

\newcommand{\regNoNeverBJS}{0.453}
\newcommand{\regNoNeverBJSse}{0.195}

\newcommand{\seratioCSApw}{1.78}
\newcommand{\seratioBJSpw}{1.31}
\newcommand{\preSigTWFEbjs}{23.8}
\newcommand{\sigflipCSAmod}{14.0}
\newcommand{\sigflipCSAtw}{7.7}
\newcommand{\corrBJSmod}{0.85}
\newcommand{\corrBJStw}{0.69}
\newcommand{\sigflipBJSmod}{7.8}
\newcommand{\sigflipBJStw}{6.8}

\newcommand{\flipCSApwpub}{27.6}
\newcommand{\flipCSApwpubAuth}{15.1}
\newcommand{\sigflipCSApwpub}{19.2}
\newcommand{\sigflipCSApwpubAuth}{6.7}
\newcommand{\nXtHighcardCmp}{35}
\newcommand{\nXtHighcardPapers}{3}
\newcommand{\nXtHighcardBJS}{4}
\newcommand{\nXtHighcardBJSinc}{31}

\newcommand{\corrXtLike}{0.90}
\newcommand{\flipXtLike}{25.7}

\newcommand{\nXtAsym}{35}
\newcommand{\corrXtAsym}{0.82}
\newcommand{\flipXtAsym}{45.7}

\newcommand{\shareUnbalCSA}{84.5}
\newcommand{\shareUnbalBJS}{85.6}

\newcommand{\corrCSAlooMin}{0.48}
\newcommand{\corrCSAlooMax}{0.74}
\newcommand{\corrBJSlooMin}{0.59}
\newcommand{\corrBJSlooMax}{0.85}

\newcommand{\flipCSAlooMin}{21.3}
\newcommand{\flipCSAlooMax}{26.7}
\newcommand{\flipBJSlooMin}{15.5}
\newcommand{\flipBJSlooMax}{19.2}
\newcommand{\hCorrCSAlooMin}{0.55}
\newcommand{\hCorrCSAlooMax}{0.71}
\newcommand{\hCorrBJSlooMin}{0.48}
\newcommand{\hCorrBJSlooMax}{0.92}
\newcommand{\hFlipCSAlooMin}{24.1}
\newcommand{\hFlipCSAlooMax}{30.8}
\newcommand{\hFlipBJSlooMin}{12.6}
\newcommand{\hFlipBJSlooMax}{33.2}

\newcommand{\nNoPassPapers}{8}
\newcommand{\nCSAnoscreen}{350}
\newcommand{\nBJSnoscreen}{411}
\newcommand{\papersCSAnoscreen}{35}
\newcommand{\papersBJSnoscreen}{40}
\newcommand{\corrCSAnoscreen}{0.63}
\newcommand{\corrBJSnoscreen}{0.80}
\newcommand{\flipCSAnoscreen}{24.6}
\newcommand{\flipBJSnoscreen}{16.1}
\newcommand{\sigflipCSAnoscreen}{9.7}
\newcommand{\sigflipBJSnoscreen}{6.1}
\newcommand{\nBigCSA}{9}
\newcommand{\nBigBJS}{2}
\newcommand{\nBigPapersCSA}{3}
\newcommand{\nBigPapersBJS}{1}
\newcommand{\shareBigCSA}{3.2}
\newcommand{\shareBigBJS}{0.7}

\newcommand{\shareBigNoNeverCSA}{33.3}

\newcommand{\shareBigNoNeverBJStext}{all}

\newcommand{\corrNeverCSA}{-0.012}
\newcommand{\corrNeverCSAse}{0.014}
\newcommand{\corrNeverBJS}{-0.027}
\newcommand{\corrNeverBJSse}{0.013}
\newcommand{\corrYearCSA}{0.280}
\newcommand{\corrYearCSAse}{0.238}
\newcommand{\corrYearBJS}{-0.055}
\newcommand{\corrYearBJSse}{0.243}

\newcommand{\nCorrSigBJS}{1}
\newcommand{\nCorrRowsCSA}{9}
\newcommand{\nCorrRowsBJS}{9}
\newcommand{\hCorrCSAse}{0.12}
\newcommand{\hCorrBJSse}{0.29}

}{}
\IfFileExists{tables/journal_funnel_numbers.tex}{

\newcommand{\fnMSnoncand}{18}
\newcommand{\fnMSdesign}{77}

\newcommand{\fnMSsynth}{13}
\newcommand{\fnMSnotshipped}{19}
\newcommand{\fnMSnopkg}{8}
\newcommand{\fnMSusable}{37}

\newcommand{\fnMStext}{161}

\newcommand{\fnMSusablePct}{48}

\newcommand{\fnMKSCnoncand}{325}
\newcommand{\fnMKSCdesign}{11}

\newcommand{\fnMKSCsynth}{4}
\newcommand{\fnMKSCnotshipped}{3}
\newcommand{\fnMKSCnopkg}{0}
\newcommand{\fnMKSCusable}{4}

\newcommand{\fnMKSCtext}{21}

\newcommand{\fnMKSCusablePct}{36}

\newcommand{\fnJMRdesign}{4}

\newcommand{\fnJMRtext}{73}

\newcommand{\fnJMcand}{90}

\newcommand{\fnJMdesign}{2}
\newcommand{\fnJMnearmiss}{22}
\newcommand{\fnJMtext}{86}

\newcommand{\fnQMEcand}{20}

\newcommand{\fnQMEnearmiss}{3}

}{}
\IfFileExists{tables/controls_numbers.tex}{\newcommand{\nCtrlRows}{114}
\newcommand{\nCtrlPapers}{12}

\newcommand{\ctrlCorrBJS}{0.99}
\newcommand{\ctrlFlipBJS}{3.0}
\newcommand{\ctrlSigBJS}{1.0}

\newcommand{\ctrlCorrCSA}{0.80}
\newcommand{\ctrlFlipCSA}{28.9}

\newcommand{\ctrlSeCSA}{2.46}
\newcommand{\ctrlNCSA}{90}
\newcommand{\ctrlNBJS}{100}
\newcommand{\ctrlPapersCSA}{10}
\newcommand{\ctrlPapersBJS}{11}

\newcommand{\ctrlCSAcells}{8}

\newcommand{\ctrlCSAlevFail}{15}
\newcommand{\ctrlCSAlevFailPapers}{3}

}{}
\IfFileExists{tables/pubcs_numbers.tex}{\newcommand{\nPubCSprinted}{54}
\newcommand{\nPubCSrepro}{52}
}{}
\IfFileExists{tables/near_miss_numbers.tex}{\newcommand{\nNearMiss}{186}
\newcommand{\nNearMissSole}{72}
\newcommand{\nNearMissTVC}{90}

\newcommand{\nNearMissExcluded}{160}
\newcommand{\nNearMissEligExcl}{3}

\newcommand{\shareNearMiss}{14.0}

}{}
\IfFileExists{tables/characteristics_numbers.tex}{

\newcommand{\nReplFail}{54}

\newcommand{\nReplTotal}{355}

\newcommand{\nReplNoPrinted}{54}
\newcommand{\nReplPrinted}{301}
\newcommand{\nReplLabelExactPrinted}{279}

\newcommand{\nReplFailSub}{10}

\newcommand{\nReplLabelApprox}{10}
\newcommand{\nReplLabelFail}{12}

\newcommand{\nReplLabelNonExact}{22}
\newcommand{\nReplApproxSub}{15}
\newcommand{\nReplApproxSubPapers}{3}
\newcommand{\nReplApproxAltered}{3}
\newcommand{\nReplApproxAlteredPapers}{3}
\newcommand{\nReplApproxVersion}{2}

\newcommand{\nReplApproxRounding}{1}

\newcommand{\nReplApproxOther}{1}

\newcommand{\nReplStata}{30}
\newcommand{\nReplR}{8}
\newcommand{\nReplBoth}{2}

\newcommand{\nCharSpecs}{355}
\newcommand{\nCharPapers}{40}

\newcommand{\charEventShare}{26.8}

\newcommand{\adoptFirstPeriod}{2019--2022}
\newcommand{\adoptLastPeriod}{2025--2026}

\newcommand{\adoptMSlastShare}{32.6}
\newcommand{\adoptMSfirstN}{29}
\newcommand{\adoptMSlastN}{89}

\newcommand{\adoptMSN}{171}
\newcommand{\adoptMSfirstYear}{2024}

\newcommand{\adoptMKSClastShare}{26.7}
\newcommand{\adoptMKSCfirstN}{14}
\newcommand{\adoptMKSClastN}{15}

\newcommand{\adoptMKSCfirstYear}{2023}
}{}
\IfFileExists{tables/xt_check_numbers.tex}{

\newcommand{\xtTwoDigitCellsUsed}{4}
\newcommand{\xtTwoDigitCells}{722}
\newcommand{\xtTwoDigitSeRatio}{3.5}

}{}
\IfFileExists{tables_appx/results_numbers.tex}{

\newcommand{\corrCSAAppx}{0.56}
\newcommand{\corrBJSAppx}{0.72}
\newcommand{\flipCSAAppx}{25.1}
\newcommand{\flipBJSAppx}{19.2}

\newcommand{\flipCSApwAppx}{19.7}
\newcommand{\flipBJSpwAppx}{16.8}

\newcommand{\nCSAAppx}{342}
\newcommand{\nBJSAppx}{365}
\newcommand{\papersCSAAppx}{29}
\newcommand{\papersBJSAppx}{30}

\newcommand{\nAddedSpecsAppx}{76}
\newcommand{\nAddedPapersAppx}{3}
}{}
\hypersetup{colorlinks=true,linkcolor=blue,citecolor=blue,urlcolor=blue}

\title{Do Modern Difference-in-Differences Estimators Change Published Findings in Business Research?}
\author{Davide Proserpio\thanks{USC Marshall School of Business; proserpi@marshall.usc.edu. Literature screening, replication, and parts of the drafting were carried out with large language model agents (Anthropic Claude) under a fixed protocol. Every judgment call is logged in the replication repository at \url{https://github.com/dadepro/modern-did}. Journal replication packages were used with the permissions their terms require.}}
\date{\today}

\begin{document}
\maketitle
\vspace{-2.5em}

\begin{abstract}
\singlespacing\noindent Staggered-adoption difference-in-differences (DiD) designs are a workhorse of empirical research in business-school journals. The two-way fixed effects (TWFE) estimator that typically implements them can place negative weights on heterogeneous treatment effects. We ask whether negative weights change published findings by reanalyzing staggered-adoption specifications in business-school journals with journal-wide public replication archives. 
Following \citet{hull2026}, we use large language model agents to identify eligible specifications, replicate the published TWFE estimate, and re-estimate each specification with the DiD-averaging estimator of \citet{callaway2021} and the imputation estimator of \citet{borusyak2024}. We re-estimate \nMainSpecs{} specifications from \nMainPapers{} papers. Across the \nCSA{} DiD-averaging and \nBJS{} imputation comparisons that pass Hull's leverage screen, from \papersCSA{} and \papersBJS{} papers, TWFE and modern estimates have correlations of \corrCSA{} and \corrBJS{}, change sign in \flipCSA{} and \flipBJS{} percent of comparisons, and differ significantly with opposite signs in \sigflipCSA{} and \sigflipBJS{} percent. Modern standard errors are \seratioCSA{} and \seratioBJS{} times those of TWFE. Including the comparisons that fail the screen leaves these figures essentially unchanged. 
Our results suggest that TWFE and heterogeneity-robust estimators disagree in a non-negligible share of the published staggered designs we re-estimate, and more often than in economics.

\end{abstract}

\section{Introduction}

Difference-in-differences is among the most common identification strategies in empirical business research. Platform entry, policy rollouts, feature launches, and regulatory changes typically arrive in different markets at different times, so the designs are almost always staggered. The standard implementation, the two-way fixed effects (TWFE) estimator, is a regression with unit and calendar-time fixed effects and a single treatment indicator. A recent econometric literature shows that, with staggered adoption and heterogeneous treatment effects, TWFE can attach negative weights to some treatment effects, and therefore produce an estimand outside the convex hull of the underlying effects, including one of the wrong sign \citep{goodmanbacon2021,dechaisemartin2020,sun2021,callaway2021,borusyak2024}. Reviews by \citet{roth2023} and \citet{baker2026} summarize the concern and the heterogeneity-robust alternatives it prompted.

Whether negative weights change the conclusions of papers that rely on TWFE is an empirical question. \citet{hull2026} answers it for economics and finds that across 1,346 staggered-adoption specifications from 108 papers in ten general-interest economics journals, TWFE estimates and modern DiD estimates have a correlation of about 0.9, and estimates change sign in 10.8 to 14.5 percent of comparisons. A hierarchical model that accounts for estimation noise implies that a statistically significant TWFE estimate with the wrong sign arises in only 0.6 to 2.1 percent of cases. \citet{chiu2026} provide comparable evidence for political science, and \citet{baker2022} document the prevalence of staggered designs in finance and accounting. No similar analysis exists for marketing or management.

The settings studied in business-school journals differ from those in economics in ways that could make the negative-weights concern more or less severe. Treatments are often firm or platform decisions observed over short panels with few adoption cohorts, and outcomes such as sales, reviews, or engagement are measured at high frequency. Treatment effects may also be more heterogeneous across adoption cohorts, for example when early adopters of a feature differ from late adopters, which is the condition under which TWFE and heterogeneity-robust estimators diverge. At the same time, large never-treated comparison groups reduce the weight TWFE places on comparisons that use already-treated units as controls, the source of its bias under dynamic effects.

In this paper, we replicate the design of \citet{hull2026} for the two business-school journals with journal-wide public replication archives: \emph{Management Science} (all departments) from 2019, when its Code and Data Disclosure Policy took effect, and \emph{Marketing Science} from 2013, when its replication policy took effect. The \emph{Journal of Marketing Research} from 2019 is also in the search frame, but its repository is restricted by default, so it contributes to the archive audit and not to the replication sample. We extend the search frame to \emph{Information Systems Research}, \emph{Manufacturing \& Service Operations Management}, \emph{Operations Research}, \emph{Organization Science}, the \emph{Journal of Marketing}, and \emph{Quantitative Marketing and Economics}. These journals have no journal-wide public replication archive in the sample window, so we screen their candidates from the article text and report eligible designs and near-misses, but do not re-estimate their papers. We use large language model agents to screen \nScreened{} articles from the three journals, and \nScreenedExtra{} from the six others, for staggered-adoption DiD specifications. The \nScreened{} are all \nCand{} candidates from a full-text and citation search and \nScreenedNonCand{} articles outside the candidate list. The agents replicate the published TWFE estimates and re-estimate each specification with the DiD-averaging estimator of \citet{callaway2021} and the imputation estimator of \citet{borusyak2024}. We then compare estimates, characterize the ex post weights that TWFE places on treatment effects, relate sign reversals to paper and specification characteristics, and estimate a hierarchical model that separates variation in estimands from estimation noise.

We re-estimate \nMainSpecs{} specifications from \nMainPapers{} papers. Following \citet{hull2026}, we compare estimators on the \nCSA{} DiD-averaging and \nBJS{} imputation comparisons, from \papersCSA{} and \papersBJS{} papers, in which neither estimate rests on a few clusters (the leverage screen). TWFE and modern DiD estimates have correlations of \corrCSA{} (DiD-averaging) and \corrBJS{} (imputation). Estimates change sign in \flipCSA{} and \flipBJS{} percent of comparisons, and in \sigflipCSA{} and \sigflipBJS{} percent the two estimates also differ significantly. Weighting papers equally, the reversal rates are \flipCSApw{} and \flipBJSpw{} percent. Both exceed the 14.5 and 10.8 percent in the economics sample of \citet{hull2026}. Modern standard errors exceed TWFE standard errors by factors of \seratioCSA{} and \seratioBJS{} on average, and the medians are \seratioCSAmed{} and \seratioBJSmed{}.

Accounting for estimation noise does not shrink the disagreement. A hierarchical model implies correlations between TWFE and modern estimands of \hCorrCSA{} and \hCorrBJS{}, with a prior sign-reversal probability of \hFlipText{}. With about 30 papers, a single paper can move these figures. Dropping one paper at a time, the estimand correlations range from \hCorrCSAlooMin{} to \hCorrCSAlooMax{} for DiD-averaging and from \hCorrBJSlooMin{} to \hCorrBJSlooMax{} for imputation.

Disagreement is concentrated in unbalanced panels, in which units enter and exit, in designs without never-treated units, and in a few papers. Conditional on these features, the absence of never-treated units is the clearest predictor of sign reversals for imputation, and no feature predicts reversals for DiD-averaging. Time-varying controls, which the protocol of \citet{hull2026} excludes, are the most common reason a staggered design does not qualify, so we also re-estimate \nCtrlPapers{} papers whose staggered designs are reported with such controls. With the published controls in its first stage, the imputation estimator tracks TWFE closely, with a correlation of \ctrlCorrBJS{} and sign reversals in \ctrlFlipBJS{} percent of \ctrlNBJS{} comparisons.

We make four contributions. First, we provide evidence on the empirical relevance of TWFE negative weights for business-school research, using the protocol of the economics benchmark so that the two literatures can be compared directly. Second, we document the designs in which TWFE and modern estimators disagree in these journals, in particular unbalanced panels, designs without never-treated units, and panels in which the outcome records whether a unit is still active and inactive units leave the data. We also record which papers already report heterogeneity-robust estimators. Third, as a by-product, we audit the replication archives of business-school journals and report the share of packages that contain usable data and the share that contain synthetic or disguised data. Fourth, we show how large language model agents can screen more than a thousand articles and replicate hundreds of specifications under a fixed protocol, with every judgment call logged and reviewable.

\section{Sample Construction}\label{sec:sample}

\subsection{Search frame}

The search frame covers every research article published in \emph{Management Science} from January 2019, \emph{Marketing Science} from January 2013, and the \emph{Journal of Marketing Research} from January 2019, through October 2026. The start years follow each journal's replication policy, as in \citet{hull2026}. \emph{Management Science} introduced its Code and Data Disclosure Policy on June 1, 2019. Because the policy applies by submission date, some articles published in 2019 and 2020 predate it. Earlier articles have no replication packages: none of 120 \emph{Management Science} articles published between 2015 and 2018, drawn at random (30 per year), has a package on the journal's replication server. \emph{Marketing Science} has required data and estimation code upon acceptance since its January--February 2013 editorial. The \emph{Journal of Marketing Research} adopted its Policy for Research Transparency on July 1, 2023. However, its Dataverse deposits are restricted to the editors by default, so for this journal only deposits that authors chose to unrestrict are usable; in our frame that is a single package, and its data are synthetic. Before July 2023 the \emph{Journal of Marketing Research} had no archive. Of 43 full texts from 2019 to 2022 in our frame, three link author-hosted materials, none for a staggered design.

Bibliographic records come from OpenAlex. The frame contains 3,197 \emph{Management Science} articles, 907 \emph{Marketing Science} articles, and 493 \emph{Journal of Marketing Research} articles, 4,597 in total, as of November 1, 2026.

Six other journals have no comparable public archive. \emph{Information Systems Research} and \emph{Manufacturing \& Service Operations Management} have no data and code policy and post no packages. \emph{Organization Science} requires third-party archiving only for submissions received after August 1, 2025. \emph{Operations Research} posts code and data as supplementary files for a few dozen recent articles. We screened \nScreenedCandExtra{} of their \nCandExtra{} candidates from the article text under the rules of the next subsection. Among them we find \nEligibleExtra{} eligible staggered designs and \nNearMissExtra{} near-misses, i.e., papers with a staggered binary design in which no main-text specification meets the rules of the protocol (Table~\ref{tab:journalfunnel} and Appendix~\ref{app:nearmiss}), none with usable replication data. These papers could be studied only with data obtained from their authors, which we do not pursue.

Outside INFORMS, the \emph{Journal of Marketing} requires a replication packet for manuscripts submitted from January 2023 and deposits it in a Harvard Dataverse collection that is restricted by default. The collection held 35 deposits by October 2026, 21 of them listing at least one public file. We screened all \nJMdeposits{}, 18 against their articles or accepted manuscripts and three from their packages alone, and downloaded the public files of all but one, which is embargoed until 2027. None contains an eligible staggered design. We screened its \fnJMcand{} candidates from 2015 onward, \fnJMtext{} of them from the article text alone: \fnJMdesign{} have an eligible staggered design, none with usable data, and \fnJMnearmiss{} are near-misses. \emph{Quantitative Marketing and Economics} follows the Springer data-availability policy, which encourages but does not require deposits. It has no deposits in Dataverse, OSF, or Zenodo. Its \fnQMEcand{} candidates contain no eligible design and \fnQMEnearmiss{} near-misses.

\subsection{Identifying candidate papers}

Abstracts in marketing rarely name the estimation method, so abstract-based screening misses most DiD papers. We therefore combine three signals. First, we searched the full text of each journal on the publisher's platform for eleven terms: ``difference-in-differences'' and its variants, ``staggered'', ``two-way fixed effects'', and the names of the main heterogeneity-robust estimators (Callaway and Sant'Anna, Sun and Abraham, Borusyak, de Chaisemartin, Goodman-Bacon). Second, we used OpenAlex citation links to flag every article in the frame that cites one of eleven key DiD methods papers, including \citet{bertrand2004}, \citet{goodmanbacon2021}, \citet{callaway2021}, \citet{sun2021}, \citet{dechaisemartin2020}, \citet{borusyak2024}, and \citet{roth2023}. Third, we flagged abstracts mentioning DiD, staggered adoption, event studies, or natural experiments.

Table~\ref{tab:candidates} reports the resulting candidate counts. Tier 1 papers show DiD language or a DiD citation together with either ``staggered'' or a heterogeneity-robust estimator. Tier 2 papers show DiD or staggered language only, and tier 3 papers have an abstract signal only. In total, \nCand{} papers are candidates, of which \nTierOne{} are in tier 1. We screened all \nScreenedCand{}. For \nScreenedText{} of them the screen rests on the article text alone: \fnMStext{} \emph{Management Science} and \fnMKSCtext{} \emph{Marketing Science} candidates without a replication package, and all \fnJMRtext{} \emph{Journal of Marketing Research} candidates, since that journal's repository is restricted. As a check on the recall of the candidate search, we also screened \nScreenedNonCand{} articles outside the candidate list: \fnMSnoncand{} \emph{Management Science} full-text search hits missing from the OpenAlex frame and \fnMKSCnoncand{} \emph{Marketing Science} packages. This additional search adds \nEligibleNonCand{} eligible designs.

\begin{table}[!htbp]\centering
\caption{Candidate papers by journal and screening tier}\label{tab:candidates}
\begin{threeparttable}
\begin{tabular}{lcccc}
\toprule
 & Tier 1 & Tier 2 & Tier 3 & Frame \\
\midrule
\emph{Management Science}, 2019--2026 & 242 & 479 & 41 & 3,197 \\
\emph{Marketing Science}, 2013--2026 & 34 & 106 & 13 & 907 \\
\emph{Journal of Marketing Research}, 2019--2026 & 10 & 54 & 9 & 493 \\
\midrule
Total & 286 & 639 & 63 & 4,597 \\
\bottomrule
\end{tabular}

\begin{tablenotes}\footnotesize
\item Notes: Tier 1 = full-text DiD language or a citation to a DiD methods paper, together with ``staggered'' or a heterogeneity-robust estimator (full-text mention or citation). Tier 2 = DiD or staggered signal only. Tier 3 = abstract-only signal. Full-text counts come from the publishers' search engines; citation flags from OpenAlex. The frame column counts research articles in the window.
\end{tablenotes}
\end{threeparttable}
\end{table}

\subsection{Screening and eligibility}

A large language model agent screened each candidate paper, working from the article text when available and otherwise from the replication package (README, code, and the regression commands extracted from it) together with the abstract. The agent applied the eligibility rules of \citet{hull2026}, reproduced in Appendix~\ref{app:protocol}.

The paper must report, in the main text, a difference-in-differences or event-study specification with a binary treatment that switches on at unit-specific dates. Specifications that appear only in an appendix, printed or online, do not count, as in \citet{hull2026}.\footnote{Appendix~\ref{app:appendixspecs} repeats the analysis with the specifications printed in the articles' appendices added. The correlations are \corrCSAAppx{} and \corrBJSAppx{}, against \corrCSA{} and \corrBJS{}.} At least two distinct adoption dates must occur among cohorts with observed untreated and treated outcomes, and an untreated comparison unit must exist in at least one treated period. The specification must be a linear regression with unit (or group) and calendar-time fixed effects, possibly with time-invariant characteristics interacted with time effects. It may not include time-varying controls, lagged outcomes, unit trends, additional treatments, continuous treatment intensity, instrumental variables, matching, stacking, or synthetic controls. Papers reporting the \citet{callaway2021}, \citet{borusyak2024}, or Gardner two-stage estimators of such a design qualify as well, as do \citet{sun2021} event studies that coincide with the \citet{callaway2021} estimator.

For each eligible paper the agent recorded every qualifying main-text specification with its table and column, outcome, treatment definition, fixed effects, weights, clustering, sample, and reported coefficient. For some specifications it also flagged whether the specification is the paper's preferred one and whether the finding appears in the abstract or introduction. A second pass of agents audited the records against the packages and, where we had the article text, against the articles, checking treatment dates, event-time bins and reference periods, and the reference estimates. The replication logs record each disagreement and the ruling it led to.

Two features of the designs in marketing and management make the screen exclude more papers than in economics. Many papers study a single platform or policy change with one adoption date, which yields a two-group design, not a staggered one, and many of the remaining staggered designs are estimated with matching-based control groups, unit-specific trends, or nonlinear count models. The replication packages add a third obstacle. Of the \nEligible{} papers with an eligible design, \nEligSynth{} come with synthetic, disguised, or pseudo data, posted under the journals' alternative disclosure plans for proprietary data. The authors' code runs on these files but, as the READMEs or deposit descriptions of most of them state, the published estimates do not reproduce. We exclude them from the main analysis. Table~\ref{tab:funnel} reports the funnel from candidates to re-estimated papers, and Table~\ref{tab:journalfunnel} breaks it down by journal.

Data availability differs by journal: about half of the eligible \emph{Management Science} designs and a third of the \emph{Marketing Science} ones come with usable data. Of the \fnMSdesign{} \emph{Management Science} papers and \fnMKSCdesign{} \emph{Marketing Science} papers with an eligible staggered design, \fnMSusable{} (\fnMSusablePct{} percent) and \fnMKSCusable{} (\fnMKSCusablePct{} percent) come with usable data. The rest come with synthetic or disguised data (\fnMSsynth{} and \fnMKSCsynth), with no analysis data (\fnMSnotshipped{} and \fnMKSCnotshipped), or with no package at all (\fnMSnopkg{} and \fnMKSCnopkg{}). None of the \fnJMRdesign{} \emph{Journal of Marketing Research} papers with an eligible design has public data.

\begin{table}[!htbp]\centering
\caption{From candidates to replicated specifications}\label{tab:funnel}
\begin{threeparttable}
\begin{tabular}{lcccc}
\toprule
 & \emph{MS} & \emph{MKSC} & \emph{JMR} & Total \\
\midrule
Articles in search frame & 3,197 & 907 & 493 & 4,597 \\
Candidate papers (tiers 1--3) & 762 & 153 & 73 & 988 \\
Candidates with a replication package & 596 & 131 & -- & 727 \\
Candidates screened & 762 & 153 & 73 & 988 \\
\quad of which from the article text alone & 161 & 21 & 73 & 255 \\
Candidates not screened (no package or article text) & 0 & 0 & 0 & 0 \\
Candidates not assessable (wrong file) & 0 & 0 & 0 & 0 \\
Non-candidates screened & 18 & 325 & 0 & 343 \\
Papers screened & 780 & 478 & 73 & 1,331 \\
\midrule
Eligible staggered design & 77 & 11 & 4 & 92 \\
\quad package data synthetic or disguised & 13 & 4 & 1 & 18 \\
\quad analysis data not in package & 19 & 3 & 3 & 25 \\
\quad no replication package & 8 & 0 & 0 & 8 \\
\quad usable data & 37 & 4 & 0 & 41 \\
\qquad re-estimated on original data & 36 & 4 & 0 & 40 \\
\qquad not re-estimated (estimate not reproduced) & 1 & 0 & 0 & 1 \\
\bottomrule
\end{tabular}

\begin{tablenotes}\footnotesize
\item Notes: Every candidate was screened: from its replication package, with the article text or abstract, when the journal's server holds one, and otherwise from the article text alone. For \emph{JMR} the latter is every candidate, because its repository is restricted. ``Not assessable'' records point to a wrong or unreadable file. Non-candidates screened are \emph{Management Science} full-text hits missing from the OpenAlex frame and \emph{Marketing Science} packages screened beyond the candidate list. An eligible staggered design has at least one main-text specification that meets every design rule of Appendix~\ref{app:protocol}, whatever the data. The indented rows partition the eligible designs, and usable data splits into re-estimated papers and \nUsableNotReest{} paper whose package does not reproduce the published estimate. ``No replication package'' counts eligible designs screened from the article text alone; three of the \emph{Management Science} ones advertise data files that the journal's server does not return. Screening follows Appendix~\ref{app:protocol}. Counts as of \today.
\end{tablenotes}
\end{threeparttable}
\end{table}

\IfFileExists{tables/journal_funnel.tex}{
\begin{table}[!htbp]\centering
\caption{Replication funnel by journal}\label{tab:journalfunnel}
\footnotesize\setlength{\tabcolsep}{3pt}
\begin{tabular}{lrrrrrrrrr}
\toprule
 & MS & MKSC & JMR & ISR & M\&SOM & OrgSci & OR & JM & QME \\
\midrule
Articles in search frame & 3,197 & 907 & 493 & 1,088 & 1,152 & 1,092 & 1,634 & 614 & 142 \\
Candidates (DiD or staggered signal) & 762 & 153 & 73 & 225 & 107 & 151 & 18 & 90 & 20 \\
\quad of which tier 1 & 242 & 34 & 10 & 57 & 28 & 26 & 2 & 21 & 6 \\
Replication package obtained & 596 & 131 & n/a & n/a & n/a & n/a & n/a & n/a & n/a \\
Candidates without a package & 166 & 22 & -- & -- & -- & -- & -- & -- & -- \\
Papers screened & 780 & 478 & 73 & 222 & 107 & 151 & 18 & 107 & 20 \\
\quad of which non-candidates & 18 & 325 & 0 & -- & -- & -- & -- & 17 & -- \\
\quad of which from the article text alone & 161 & 21 & 73 & 222 & 107 & 151 & 18 & 86 & 20 \\
Candidates not screened & 0 & 0 & 0 & 3 & 0 & 0 & 0 & 0 & 0 \\
Eligible staggered design & 77 & 11 & 4 & 14 & 5 & 8 & 0 & 2 & 0 \\
\quad synthetic or disguised data & 13 & 4 & 1 & -- & -- & -- & -- & 1 & -- \\
\quad analysis data not shipped & 19 & 3 & 3 & -- & -- & -- & -- & -- & -- \\
\quad no replication package & 8 & 0 & 0 & 14 & 5 & 8 & 0 & 1 & 0 \\
\quad usable data & 37 & 4 & 0 & -- & -- & -- & -- & -- & -- \\
\qquad re-estimated on original data & 36 & 4 & 0 & -- & -- & -- & -- & -- & -- \\
\qquad not re-estimated & 1 & 0 & 0 & -- & -- & -- & -- & -- & -- \\
\midrule
Near-miss (no qualifying specification) & 153 & 31 & 2 & 65 & 25 & 44 & 0 & 22 & 3 \\
\midrule
Usable data / eligible design & 48 & 36 & 0 & -- & -- & -- & -- & -- & -- \\
Re-estimated / candidates & 5 & 3 & 0 & -- & -- & -- & -- & -- & -- \\
\bottomrule
\end{tabular}

\begin{minipage}{0.95\textwidth}\footnotesize\vspace{2pt}
Notes: Counts as of \today. Candidates are articles whose full text or abstract contains DiD, staggered-adoption, or modern-estimator language, or that cite one of eleven method papers. Tier 1 adds a staggered or modern-estimator signal. ``Eligible staggered design'' counts papers with at least one main-text specification meeting the protocol's design rules regardless of data availability. ``Usable data'' requires the package to contain the original or an authors-transformed version of the analysis data. ISR, M\&SOM, Organization Science, Operations Research, JM, and QME have no journal-wide public replication archive in the sample window; we screened their candidates from the article text, so they contribute eligible designs and near-misses but no replications. The three candidates not screened are 2018 ISR articles whose full text we could not obtain. JMR's repository is restricted by default, so it contributes candidates and screening records but no replications. Near-misses have a staggered binary design in which no main-text specification qualifies (Appendix~\ref{app:nearmiss}). The 17 JM non-candidates are Dataverse deposits screened from their packages (Section~\ref{sec:sample}).
\end{minipage}
\end{table}}{}

\subsection{Replication}

We requested the replication packages of \emph{Management Science} and \emph{Marketing Science} articles through the journals' replication-file form, for every candidate and for the articles outside the candidate list described above. The server returned \nPkgFiles{} files, \nCandPkg{} of them for candidate papers and \nPkgNonCand{} for the others. It holds no file for most articles published before the policies took effect. The packages are written in Stata for \nReplStata{} papers, in R for \nReplR{}, and in both for \nReplBoth{}. For each eligible paper, a replication agent located the analysis dataset and the code producing the published columns and wrote a machine-readable description of each specification (data file, unit and time identifiers, treatment indicator, weights, clustering, and time-invariant covariates) that drives the re-estimation. Replication then proceeds in two steps. First, the agent runs the authors' code and compares its output with the printed estimate. We classify this replication as exact, approximate (same sign, within 10 percent), or failed, and document deviations in our replication logs. 

Of the \nReplTotal{} specifications in the main sample, \nReplNoPrinted{} have no printed value, because the estimate is shown only in a figure or, in one case, because the printed column adds controls that the specification omits. For these the agent checks the authors' code against the output it stores. Of the \nReplPrinted{} with a printed value, the authors' code reproduces it exactly for \nReplLabelExactPrinted{}. The other \nReplLabelNonExact{} fall into five groups. The package contains a random subsample of the analysis data, posted for confidentiality, for \nReplApproxSub{} specifications from \nReplApproxSubPapers{} papers, and a disclosure-reduced or transformed version of the data for \nReplApproxAltered{} specifications from \nReplApproxAlteredPapers{} papers. In one paper, \nReplApproxVersion{} estimates come from a different version of the \texttt{did} package. For \nReplApproxRounding{} the printed coefficient is truncated rather than rounded, and for \nReplApproxOther{} it contradicts its own t-statistic. Of these \nReplLabelNonExact{}, \nReplLabelApprox{} are within 10 percent of the printed value and \nReplLabelFail{} differ by more or in sign.

Second, the pipeline re-estimates each specification under the common protocol of Section~\ref{sec:methods} and Appendix~\ref{app:protocol}, which follows \citet{hull2026}: one binary treatment that switches on once, the authors' fixed effects, weights, and clustering, time-invariant covariates only, and an outcome standardized in the regression sample. The protocol departs from the authors' regression only where one of its rules requires it, so that TWFE and the two modern estimators share one sample and one treatment definition. Examples are dropping post-adoption observations in which the treatment has switched off, assigning units treated before the sample starts their true adoption date, and holding a time-varying industry code fixed. The protocol's estimate differs from the printed one by more than 10 percent or in sign for \nReplFail{} specifications, \nReplFailSub{} of which belong to a single paper whose package contains a random subsample of the data. These are the cases in which the protocol departs from the authors' regression, the package contains a random subsample of the data, the package data are transformed, or a printed value is in error.

Table~\ref{tab:characteristics} describes the \nCharSpecs{} specifications from \nCharPapers{} papers, in the format of \citet{hull2026}. Nearly all come from papers published in 2021 or later, about one in six from \emph{Marketing Science}, and about one in three from papers that themselves report a heterogeneity-robust estimator. Within \emph{Management Science}, accounting, strategy, and information systems contribute most specifications, and marketing about one in nine. Event studies are \charEventShare{} percent of specifications. Never-treated units account for \shareNeverMean{} percent of observations on average, against 41.5 percent in the economics sample, and always-treated units for \shareAlwaysMean{} percent, against 6.9 percent there.

\begin{table}[!htbp]\centering
\caption{Specification characteristics}\label{tab:characteristics}
\begin{threeparttable}
\small
\begin{tabular}{lcc}
\toprule
 & All specifications & Pass leverage screen \\
\midrule
Publication year & 2023.7 & 2023.7 \\
Published in 2021 or later & 0.946 & 0.938 \\
\emph{Marketing Science} & 0.166 & 0.192 \\
Paper citations (mean) & 122.1 & 133.4 \\
Paper citations (median) & 21 & 35 \\
Paper reports a modern estimator & 0.324 & 0.352 \\
Result in abstract or introduction & 0.259 & 0.215 \\
Preferred specification & 0.177 & 0.173 \\
Event study & 0.268 & 0.306 \\
Adoption cohorts (median) & 13 & 19 \\
Always-treated share (static specifications) & 0.044 & 0.037 \\
Never-treated share & 0.383 & 0.366 \\
\midrule
\multicolumn{3}{l}{\emph{Management Science department (share of MS specifications)}} \\
\quad Accounting & 0.328 & 0.298 \\
\quad Finance & 0.122 & 0.089 \\
\quad Information systems & 0.169 & 0.202 \\
\quad Strategy, innovation and organizations & 0.216 & 0.218 \\
\quad Marketing & 0.111 & 0.133 \\
\quad Operations and other & 0.054 & 0.060 \\
\midrule
Specifications & 355 & 307 \\
Papers & 40 & 32 \\
\bottomrule
\end{tabular}

\begin{tablenotes}\footnotesize
\item Notes: Means or shares over the specifications of the main sample (static estimates and event studies, each counted once), for all specifications and for those passing the leverage screen (Section~\ref{sec:methods}). Citations are OpenAlex counts of the paper. ``Reports a modern estimator'' means the paper reports a Callaway--Sant'Anna, Sun--Abraham, Borusyak--Jaravel--Spiess, Gardner, or de Chaisemartin--D'Haultf\oe{}uille estimate. The department is the one named in the editor's acceptance line of the article. The always-treated share is defined for static specifications and the never-treated share in the regression sample of each specification.
\end{tablenotes}
\end{threeparttable}
\end{table}

\section{TWFE and Modern DiD}\label{sec:methods}

For each specification, the TWFE estimate $\hat\beta^{TWFE}$ comes from
\begin{equation}
Y_j = \alpha_{u(j)} + \lambda_{t(j)} + D_j'\beta^{TWFE} + W_{u(j)}'\gamma_{t(j)} + \varepsilon_j,
\end{equation}
where $u(j)$ indexes the unit or group, $t(j)$ calendar time, $D_j$ the treatment indicator (or a vector of relative-time indicators in event studies), and $W_{u(j)}$ fixed unit characteristics that may be interacted with time effects. Under no anticipation and parallel trends, the static TWFE estimand is $\beta^{TWFE} = E[\psi_j\beta_j]/E[\psi_j]$ with ex post weights $\psi_j = \tilde D_j D_j$, where $\tilde D_j$ is the residual from projecting $D_j$ on the fixed effects \citep{borusyak2024aea}. Weights can be negative when $\tilde D_j<0$ for treated observations.

The DiD-averaging estimator follows \citet{callaway2021}. Cohort-by-period average treatment effects on the treated are estimated as DiD contrasts relative to the last pre-adoption period, using units not yet treated at $t$ as controls (including never-treated units) and adjusting linearly for $W_{u(j)}$. They are then aggregated with convex weights proportional to the number of treated observations. We apply these settings to every specification, including the \nPubCSspecs{} specifications of the \nPubCSpapers{} papers that themselves publish a Callaway--Sant'Anna estimate. As a check on the replication, we also estimate these specifications with the authors' implementation: their comparison group, their construction of the cohort-by-period cells, their aggregation, and their cohort weights. That estimate matches the published one to the last printed decimal in \nPubCSrepro{} of the \nPubCSprinted{} specifications with a printed estimate (Appendix Table~\ref{tab:pubcs}), and Section~\ref{sec:results} reports how the main results change when it replaces Hull's settings for these papers.

The imputation estimator follows \citet{borusyak2024}: the fixed-effects model is fit on untreated observations only, counterfactual outcomes are imputed for treated observations, and treatment effects are averaged. We compute it with the two-stage procedure of \citet{gardner2026}, which accommodates repeated cross-sections. Its point estimate coincides with the \citet{borusyak2024} imputation estimator when unit fixed effects are estimated from all untreated observations, including units with a single pre-period. We verified this equality against a direct implementation. The standard errors are those of \citet{gardner2026}.

Event studies keep their published bins and reference period \citep{hull2026}. The TWFE regression uses the authors' relative-time indicators. The modern estimators average their cohort-by-period effects within the same bins and subtract the estimated effect in the reference period, which is zero by construction when the reference is the last pre-adoption period alone. We summarize each event study by separate averages of the post- and pre-treatment bins over common supported horizons, following \citet{hull2026}. For the imputation estimator, when the reference period is the last pre-adoption period alone, the pre-treatment average follows \citet{borusyak2024}: we regress the outcome on the fixed effects and lead indicators using untreated observations and average the lead coefficients over the pre-treatment bins. When the authors use another reference set, the pre-treatment average is the explicit-reference contrast of \citet{hull2026}: the mean of the imputation residuals in the pre-treatment bins minus their mean in the reference set.

Outcomes are normalized to unit standard deviation in the regression sample. Adoption dates come from the original treatment history \citep{hull2026}: units that adopt before the first sample period keep their true adoption date, which sets their TWFE event time, and are always-treated for the modern estimators. Data with several observations per unit or group and period are repeated cross-sections in the sense of \citet{hull2026}, and the DiD-averaging estimator then uses its repeated cross-section form.

Time-invariant characteristics enter the DiD-averaging estimator as category indicators. Characteristics with more than 60 categories, mostly four-digit industry codes, cannot enter this way. With 260 industry indicators, the cohort-by-period cell regressions are rank-deficient in all or nearly all cells and the estimator returns no estimate. A two-digit classification returns an estimate from only \xtTwoDigitCellsUsed{} of \xtTwoDigitCells{} cells, with standard errors \xtTwoDigitSeRatio{} times those of the estimator without the characteristic. Imputation, which fits one pooled fixed-effects regression, can include them, although with industry-by-year effects and small cohorts most of these specifications fail the leverage screen described below, and \nXtHighcardBJS{} enter the imputation comparison. For specifications with such a characteristic we therefore report no DiD-averaging estimate, and TWFE and imputation keep the authors' regression. This costs \nXtHighcardCmp{} DiD-averaging comparisons from \nXtHighcardPapers{} papers. Our choice departs from \citet{hull2026}, who adjusts for $W_{u(j)}$ throughout, and from the alternative of dropping the characteristic from DiD-averaging alone, which compares estimators that adjust for different covariates. That alternative would attribute the covariate adjustment to the estimator. On the \nXtAsym{} comparisons with such a characteristic, TWFE with the industry-by-year effects and DiD-averaging without them have a correlation of \corrXtAsym{} and change sign in \flipXtAsym{} percent of cases. When TWFE is also re-estimated without the characteristic, so that both estimators adjust for the same covariates, its correlation with the same DiD-averaging estimates is \corrXtLike{} and the sign changes in \flipXtLike{} percent. 

In total, the DiD-averaging estimator has no estimate in \nCSAunsupRowsAll{} comparisons from \nCSAunsupPapersAll{} papers, or \nCSAunsupRows{} comparisons from \nCSAunsupPapers{} papers after the leverage screen described below, the high-cardinality cases included. The remaining cases have two sources. In \nCSArankdefPapers{} papers, the authors' time-invariant characteristics leave the cohort-by-period regressions rank-deficient, and in \nCSAnocellsPapers{} the \texttt{did} package finds no valid cohort-by-period comparison. These comparisons contribute TWFE and, where the screen admits it, imputation estimates only. As for the high-cardinality cases, we do not drop the characteristics to obtain a DiD-averaging estimate.

To test whether two estimates differ, we use their covariance from a joint weighted (Bayesian) bootstrap with 999 draws, as in \citet{hull2026}. Each draw multiplies the analysis weights by independent unit-exponential weights, drawn at the level of the clustering variable, and re-estimates TWFE and both modern estimators on the same draw. The DiD-averaging estimator is re-estimated in each draw from exact cell-level implementations that reproduce the \texttt{did} package to numerical precision. Where it adjusts for covariates (the authors' time-invariant covariates interacted with time, or the time-varying controls of Section~\ref{sec:controls}), or where the cell-level replica does not reproduce the package estimate, the bootstrap omits the DiD-averaging estimator and the covariance is set to zero. In the main sample, \nCovCSA{} of \nCSA{} DiD-averaging comparisons and \nCovBJS{} of \nBJS{} imputation comparisons have a bootstrap covariance. In \nNegVarCSA{} DiD-averaging and \nNegVarBJS{} imputation comparisons, the analytic standard errors and the bootstrap covariance imply a non-positive variance for the difference, because the analytic standard errors fall slightly below the bootstrap standard deviations while the bootstrap correlation is close to one. \citet{hull2026} does not address this case. For these comparisons we scale the bootstrap correlation by the analytic standard errors, which always yields a valid variance. The main test combines the analytic standard errors with the bootstrap covariance, consistent with \citet{hull2026}, who obtains the covariances from the bootstrap. As a robustness check, we report the test with the variance taken entirely from the bootstrap draws.

The leverage screen of \citet{hull2026} retains a pair of estimates when both estimators have at least 10 effective clusters and a maximum cluster share of at most 25 percent. Each estimator's effective number of clusters and maximum cluster share come from its weights on the outcome, holding its cohort shares, cell counts, and analysis weights fixed. For DiD-averaging with covariates, the weights are those of its cell-level regressions. Where the untreated observations do not identify the imputed counterfactual of some treated observations, so that the imputation estimate depends on how the fixed effects are normalized, the imputation comparison fails the screen (\nBJSnotIdent{} comparisons, \nXtHighcardBJSinc{} of them specifications with a high-cardinality characteristic). Where the problem is too large for the \texttt{did} package (more than 4,000 cohort-by-period cells or 3 million observations, or an influence function that does not fit in memory), we compute the same estimator directly from cohort and comparison-group means, with bootstrap standard errors (\nCSAlean{} comparisons). Where it also adjusts for covariates, which rules out this shortcut, we compute it cell by cell from the same doubly robust or regression cell estimates, accumulating the influence function of the reported summary as a single vector (\ctrlCSAcells{} comparisons, all in Section~\ref{sec:controls}). Both implementations reproduce the package's estimates and standard errors to numerical precision wherever the package runs.

Unbalanced panels need particular care in this literature, because many firm, store, and household panels lose and gain units over time. Following \citet{hull2026}, the DiD-averaging estimator compares cohort and comparison-group sample averages in period $t$ and in the base period $g-1$, using whichever units are observed in each period. We implement it with the unbalanced-panel option of the \texttt{did} package, which treats each cohort-by-period cell as a repeated cross-section. Software defaults differ on this point. By default, the \texttt{did} package drops every unit that is not observed in all periods, whereas Stata's \texttt{csdid} estimates each cohort-by-period contrast on the units observed in both the base period and period $t$. In the most unbalanced panels in our sample, the three choices give different answers on the same data. In one panel, no treated cohort survives the \texttt{did} default, because no treated unit is observed in its base period. In another, with the same never-treated comparison group, the cohort-average estimate and the \texttt{csdid} estimate, which uses only units observed in both periods of each contrast, have opposite signs. TWFE and imputation instead compare each unit with itself through unit fixed effects, so when units enter or exit, DiD-averaging can differ from the other two estimators for that reason alone.

\section{Results}\label{sec:results}

\subsection{Comparisons}

\IfFileExists{tables/comparison.tex}{
\begin{table}[!htbp]\centering
\caption{TWFE versus modern DiD estimates}\label{tab:comparison}
\begin{threeparttable}
\begin{tabular}{lcc}
\toprule
 & DiD-averaging & Imputation \\
\midrule
Correlation of estimates & 0.607 & 0.705 \\
Share of sign reversals & 0.251 & 0.180 \\
Share of significant sign reversals & 0.102 & 0.072 \\
Mean SE ratio (modern / TWFE) & 1.61 & 1.30 \\
\midrule
\multicolumn{3}{l}{\emph{Paper-weighted (each paper counts once)}} \\
Correlation of estimates & 0.621 & 0.350 \\
Share of sign reversals & 0.199 & 0.164 \\
Share of significant sign reversals & 0.112 & 0.115 \\
\midrule
\multicolumn{3}{l}{\emph{Excluding pre-window averages of event studies}} \\
Correlation of estimates & 0.643 & 0.704 \\
Share of sign reversals & 0.264 & 0.167 \\
Share of significant sign reversals & 0.096 & 0.061 \\
Comparisons & 239 & 264 \\
\midrule
Comparisons & 283 & 306 \\
Papers & 28 & 29 \\
\bottomrule
\end{tabular}

\begin{tablenotes}\footnotesize
\item Notes: Each comparison pairs a TWFE estimate with the DiD-averaging (Callaway--Sant'Anna) or imputation (Borusyak--Jaravel--Spiess) estimate of the same specification, both in units of the regression-sample outcome standard deviation. Sign reversals are pairs with opposite signs. Significant reversals are sign reversals whose estimate difference is also significant at the 5 percent level, using the analytic standard errors and the covariance of the two estimates from a joint 999-draw cluster Bayesian bootstrap. Where these imply a non-positive variance for the difference, the bootstrap correlation is scaled by the analytic standard errors (Section~\ref{sec:methods}). The sample is restricted to pairs for which both estimators have at least 10 effective clusters and a maximum cluster share of at most 25 percent.
\end{tablenotes}
\end{threeparttable}
\end{table}
}{\todo{Table: comparison summary (generated by code/R/analysis.R).}}

\IfFileExists{figures/fig2.pdf}{
\begin{figure}[!htbp]\centering
\caption{Modern difference-in-differences versus two-way fixed effects}\label{fig:compare}
\includegraphics[width=\textwidth]{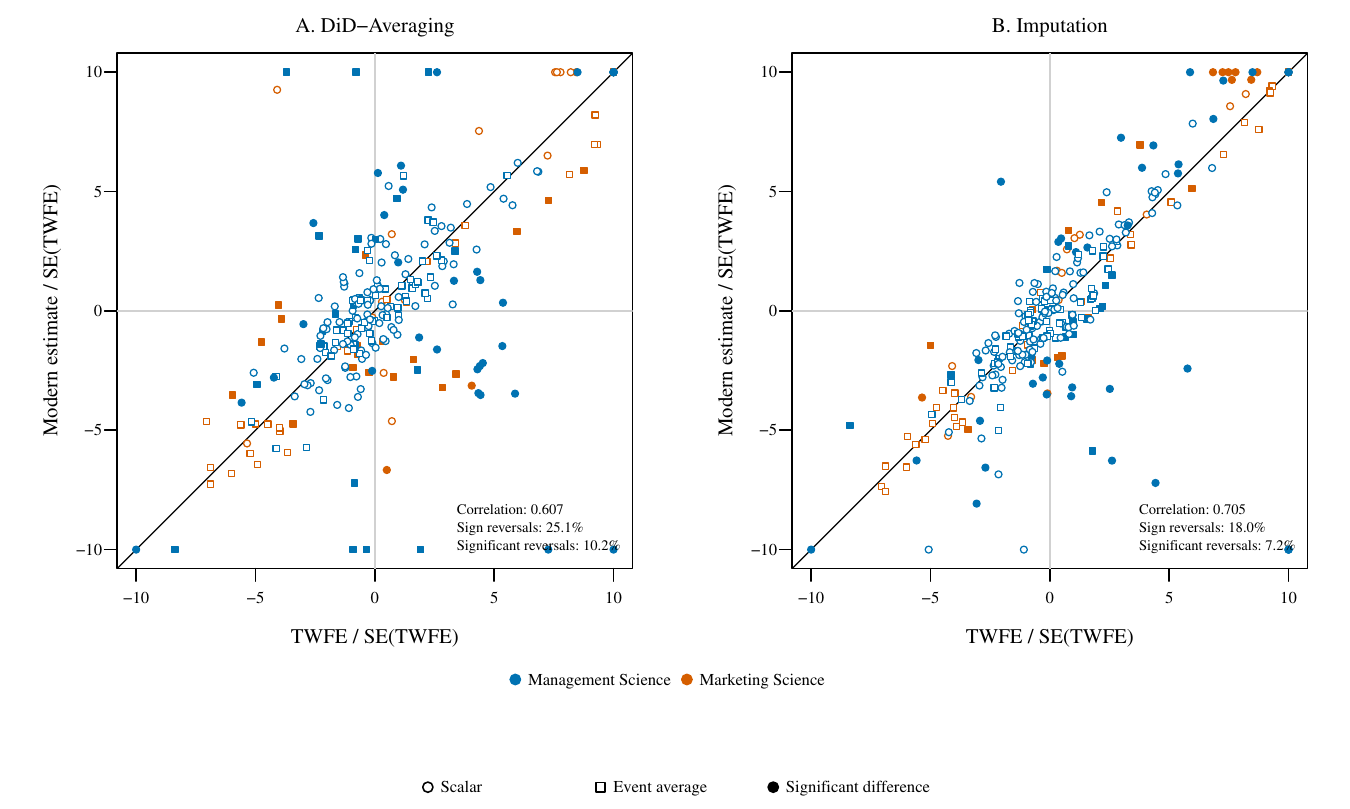}
\begin{minipage}{0.95\textwidth}\footnotesize\vspace{2pt}
Notes: This figure compares DiD-averaging estimates \citep{callaway2021} and imputation estimates \citep{borusyak2024} with TWFE estimates for the specifications in the main sample. Each point is a scalar estimate (circle) or an event-study window average (square), with separate pre- and post-window averages where available. Both coordinates are divided by the TWFE standard error, and the diagonal denotes equal estimates. Filled symbols indicate that the difference between the two estimates is statistically significant at the 5 percent level. Coordinates are capped at $\pm 10$, but uncapped values are used for the statistics. ``Sign reversals'' is the share of pairs with opposite signs, and ``Significant reversals'' is the share with a sign reversal and a significant difference.
\end{minipage}
\end{figure}
}{}

Table~\ref{tab:comparison} summarizes the \nCSA{} DiD-averaging and \nBJS{} imputation comparisons that pass the leverage screen (Section~\ref{sec:methods}), from \papersCSA{} and \papersBJS{} papers. They come from \nMainSpecs{} re-estimated specifications from \nMainPapers{} papers with original data, with the pre- and post-window averages of each event study counted as separate comparisons (\nMainComparisons{} in total). The screen removes every specification of \nNoPassPapers{} papers, in which treatment varies across a few states, adoption cohorts, or other clusters.
The correlation between TWFE and modern estimates is \corrCSA{} for DiD-averaging and \corrBJS{} for imputation (Figure~\ref{fig:compare} plots the estimates). Estimates change sign in \flipCSA{} percent of DiD-averaging comparisons and \flipBJS{} percent of imputation comparisons. Reversals accompanied by a statistically significant difference between the two estimates are rarer, at \sigflipCSA{} and \sigflipBJS{} percent. The economics benchmarks in \citet{hull2026} are correlations of 0.88 and 0.93, reversal rates of 14.5 and 10.8 percent, and significant reversal rates of 3.5 and 2.2 percent. Our correlations are lower and our reversals more frequent. 

Without the leverage screen, the comparisons cover \papersCSAnoscreen{} papers for DiD-averaging and \papersBJSnoscreen{} for imputation, instead of \papersCSA{} and \papersBJS{}, and the correlations are \corrCSAnoscreen{} and \corrBJSnoscreen{} against \corrCSA{} and \corrBJS{} with the screen (Appendix~\ref{app:noscreen}), so the screen does not drive the low correlations. 

Because a few papers contribute many specifications, the middle panel of Table~\ref{tab:comparison} reweights the comparisons so that each paper counts once. The paper-weighted reversal shares are \flipCSApw{} and \flipBJSpw{} percent, and the paper-weighted correlations are \corrCSApw{} and \corrBJSpw{}. With \papersCSA{} and \papersBJS{} papers, a single paper can move these statistics. Dropping one paper at a time, the correlation ranges from \corrCSAlooMin{} to \corrCSAlooMax{} for DiD-averaging and from \corrBJSlooMin{} to \corrBJSlooMax{} for imputation, and the shares of sign reversals stay between \flipCSAlooMin{} and \flipCSAlooMax{} percent and between \flipBJSlooMin{} and \flipBJSlooMax{} percent. Appendix Table~\ref{tab:loo} reports the range of every statistic.

The significance test uses the variance of the difference between the TWFE and the modern estimate, built from the analytic standard error of each estimate and the bootstrap covariance between the two, as in \citet{hull2026}. Because both estimators are computed on the same sample, their estimates are positively correlated across bootstrap draws, with median correlations of \bootCorCSA{} and \bootCorBJS{}. This covariance lowers the variance of the difference, so allowing for it raises the share of significant reversals. With the covariance set to zero, the shares are \sigflipCSAzero{} and \sigflipBJSzero{} percent instead of \sigflipCSA{} and \sigflipBJS{}. The conclusion does not depend on how the variance of the difference is constructed. Scaling the bootstrap correlation by the analytic standard errors gives \sigflipCSAcorr{} and \sigflipBJScorr{} percent, and taking the variance entirely from the bootstrap draws gives \sigflipCSAbvar{} and \sigflipBJSbvar{} percent. Counting all significant differences regardless of sign, the shares are \sigdiffCSA{} and \sigdiffBJS{} percent, or \sigdiffCSAzero{} and \sigdiffBJSzero{} percent with zero covariance.

Disagreement is concentrated in unbalanced panels, in which units enter or leave the data. These panels account for \shareUnbalCSA{} percent of DiD-averaging and \shareUnbalBJS{} percent of imputation comparisons. In balanced panels and repeated cross-sections, the correlations between TWFE and modern estimates are \corrCSAbal{} for DiD-averaging and \corrBJSbal{} for imputation, close to the economics benchmarks. In unbalanced panels they are \corrCSAunbal{} and \corrBJSunbal{}. For DiD-averaging, entry and exit can contribute to this gap directly. Following \citet{hull2026}, for unbalanced panels, its cohort averages use whichever units are observed in each period, whereas TWFE and imputation compare each unit with itself through unit fixed effects \citep[see][on DiD when the composition of the sample changes over time]{santanna2023}. Despite Imputation and TWFE having the same averaging procedure, Imputation also agrees less with TWFE in unbalanced panels, so imbalance likely stands in for other features of these designs, such as entry and exit that depends on the outcome and the lack of never-treated units (Section~\ref{sec:vshull}).

For the \nPubCSpapers{} papers that publish a Callaway--Sant'Anna estimate, the main DiD-averaging estimates use the settings of \citet{hull2026}, as for every other specification, although the authors' own settings differ (Section~\ref{sec:methods}).\footnote{Appendix Table~\ref{tab:authcs} replaces Hull's settings with the authors' for these papers.} Replacing Hull's settings with the authors' for these papers changes little across all \nCSA{} DiD-averaging comparisons: the shares of sign reversals and significant reversals fall from \flipCSA{} and \sigflipCSA{} percent to \flipCSAauth{} and \sigflipCSAauth{} percent, and the correlation with TWFE is \corrCSAauth{}. Within the \nPubCSspecs{} specifications of the papers that publish a Callaway--Sant'Anna estimate, the change is larger when each paper counts once: the paper-weighted shares fall from \flipCSApwpub{} to \flipCSApwpubAuth{} percent and from \sigflipCSApwpub{} to \sigflipCSApwpubAuth{} percent. Across these specifications, the median estimate moves by \authMedGapSE{} standard errors, and \nAuthMoveOneSE{} estimates from \nAuthMoveOneSEpapers{} papers move by more than one standard error.
Following \citet{hull2026}, event studies enter the main comparison through both their post-window and pre-window averages, each as a separate comparison point. Pre-window averages are expected to be near zero under parallel trends, so sign reversals between estimators carry less information there. The lower panel of Table~\ref{tab:comparison} therefore drops the pre-window averages, as \citet{hull2026} does in a robustness check. The reversal rates become \flipCSApost{} and \flipBJSpost{} percent, against 12.8 and 8.3 percent in his sample. Table~\ref{tab:pretrends} examines the pre-window averages of the \nPreCSA{} and \nPreBJS{} event studies in the DiD-averaging and imputation samples on their own, as pre-trend summaries. The TWFE pre-average is significant at the 5 percent level in \preSigTWFE{} percent of the DiD-averaging sample and \preSigTWFEbjs{} percent of the imputation sample, against \preSigCSA{} percent for DiD-averaging and \preSigBJS{} percent for imputation.

\IfFileExists{tables/pretrends.tex}{
\begin{table}[!htbp]\centering
\caption{Pre-window averages in event studies}\label{tab:pretrends}
\small
\begin{tabular}{lcc}
\toprule
 & DiD-averaging & Imputation \\
\midrule
Correlation of pre-window averages & 0.330 & 0.814 \\
Share of sign reversals & 0.182 & 0.262 \\
Share of significant sign reversals & 0.136 & 0.143 \\
Share of TWFE pre-averages significant at 5\% & 0.273 & 0.238 \\
Share of modern pre-averages significant at 5\% & 0.250 & 0.238 \\
\midrule
Event studies & 44 & 42 \\
Papers & 9 & 10 \\
\bottomrule
\end{tabular}

\begin{minipage}{0.85\textwidth}\footnotesize\vspace{2pt}
Notes: One row per event-study specification passing the leverage screen. The pre-window average is the mean of the relative-time coefficients over the published pre-treatment bins, relative to the published reference period. The Callaway--Sant'Anna version uses the universal base period, and the imputation version averages the lead coefficients of a regression estimated on untreated observations \citep{borusyak2024} when the reference period is the last pre-adoption period alone, and otherwise the explicit-reference contrast of \citet{hull2026} (Section~\ref{sec:methods}).
\end{minipage}
\end{table}}{}

Papers that themselves report a modern estimator enter the comparison like all others. Where a specification is reported only as a modern estimate, we construct its TWFE counterpart. Splitting the comparisons by whether the paper reports a modern estimator tests for selection into modern methods: if authors adopt them when TWFE gives a different answer, reversals should be more common among papers that report one. Table~\ref{tab:byreported} reports the split. In the \nTwPapers{} papers that report TWFE only, estimates reverse sign in \flipCSAtw{} percent of DiD-averaging comparisons, and the correlation is \corrCSAtw{}. In the \nModPapers{} papers that report a modern estimator, the reversal rate is \flipCSAmod{} percent and the correlation \corrCSAmod{}. Therefore, reversals are slightly less common in papers that report a modern estimator, as in \citet{hull2026}, for both estimators. 
Significant reversals, however, are more common in these papers, at \sigflipCSAmod{} against \sigflipCSAtw{} percent for DiD-averaging and \sigflipBJSmod{} against \sigflipBJStw{} percent for imputation. The correlation is lower in these papers for DiD-averaging, at \corrCSAmod{} against \corrCSAtw{}, and higher for imputation, at \corrBJSmod{} against \corrBJStw{}.

\IfFileExists{tables/comparison_by_reported.tex}{
\begin{table}[!htbp]\centering
\caption{TWFE versus modern DiD estimates, by the estimator the paper reports}\label{tab:byreported}
\small
\begin{tabular}{lcccc}
\toprule
 & \multicolumn{2}{c}{DiD-averaging} & \multicolumn{2}{c}{Imputation} \\
\cmidrule(lr){2-3} \cmidrule(lr){4-5}
Paper reports & TWFE only & modern DiD & TWFE only & modern DiD \\
\midrule
Correlation of estimates & 0.739 & 0.414 & 0.685 & 0.846 \\
Share of sign reversals & 0.254 & 0.246 & 0.195 & 0.155 \\
Share of significant sign reversals & 0.077 & 0.140 & 0.068 & 0.078 \\
Mean SE ratio (modern / TWFE) & 1.61 & 1.60 & 1.36 & 1.22 \\
\midrule
Comparisons & 169 & 114 & 190 & 116 \\
Papers & 19 & 9 & 20 & 9 \\
\bottomrule
\end{tabular}

\begin{minipage}{0.9\textwidth}\footnotesize\vspace{2pt}
Notes: Same sample and statistics as Table~\ref{tab:comparison}. ``Modern DiD'' papers report at least one Callaway--Sant'Anna, Sun--Abraham, Borusyak--Jaravel--Spiess, Gardner, or de Chaisemartin--D'Haultf\oe{}uille estimate. Where a specification is reported only as a modern estimate, its TWFE counterpart is constructed from the same specification.
\end{minipage}
\end{table}}{}

Switching estimators costs precision. Standard errors of the DiD-averaging estimator are on average \seratioCSA{} times the TWFE standard error, and those of the imputation estimator \seratioBJS{} times, compared with 1.47 and 1.14 in \citet{hull2026}. The median ratios are \seratioCSAmed{} and \seratioBJSmed{}, and the paper-weighted means \seratioCSApw{} and \seratioBJSpw{}. Our imputation standard errors are those of \citet{gardner2026}. \citet{hull2026} does not state which standard errors he uses for this estimator, so the comparison of imputation standard errors is approximate. For DiD-averaging, part of the cost comes from unbalanced panels, where cohort averages use only the units observed in each period, and from the many small cohorts, whose cohort-by-period effects are imprecisely estimated. The median DiD-averaging comparison that passes the screen has \medianCohorts{} adoption cohorts, fewer than the specifications in Table~\ref{tab:characteristics}.

Negative ex post weights carry more mass than in economics. The median share of treated observations with a negative weight is \negRowsMedian{} percent and the median negative share of absolute weight is \negAbsMedian{} percent (Table~\ref{tab:negweights}), against 8 and 1.4 percent in \citet{hull2026}. The mean shares of never-treated and always-treated observations are \shareNeverMean{} and \shareAlwaysMean{} percent, close to or below the 41.5 and 6.9 percent in his sample, so the larger negative weights do not come from fewer never-treated or more always-treated units.

\IfFileExists{tables/negweights.tex}{
\begin{table}[!htbp]\centering
\caption{Negative ex post weights in static TWFE specifications}\label{tab:negweights}
\begin{threeparttable}
\begin{tabular}{lcccc}
\toprule
 & Mean & Median & SD & N \\
\midrule
Share of treated rows with negative weight & 0.219 & 0.224 & 0.153 & 260 \\
Negative share of absolute weight & 0.108 & 0.054 & 0.108 & 260 \\
\bottomrule
\end{tabular}

\begin{tablenotes}\footnotesize
\item Notes: One observation per static TWFE specification with original data, before the leverage screen. The ex post weights follow \citet{borusyak2024aea}.
\end{tablenotes}
\end{threeparttable}
\end{table}
}{}

Sharp disagreements are rare and concentrated. Among the comparisons that pass the screen, the two estimates differ by more than half an outcome standard deviation in \nBigCSA{} DiD-averaging and \nBigBJS{} imputation comparisons, from \nBigPapersCSA{} and \nBigPapersBJS{} papers, which is \shareBigCSA{} and \shareBigBJS{} percent of the comparisons. All of them come from unbalanced panels, and \shareBigNoNeverCSA{} percent of the DiD-averaging and \shareBigNoNeverBJStext{} of the imputation cases come from designs without never-treated units. 

\subsection{Correlates}
Table~\ref{tab:correlates} relates sign reversals to specification and paper characteristics, as in \citet{hull2026}. Few characteristics predict reversals with any precision: none of the \nCorrRowsCSA{} coefficients for DiD-averaging and only \nCorrSigBJS{} of the \nCorrRowsBJS{} for imputation are more than twice their standard errors. The clearest association is the one \citet{hull2026} finds. For imputation, ten percentage points more never-treated observations are associated with a change of \corrNeverBJS{} (standard error \corrNeverBJSse{}) in the probability of a sign reversal, against a reduction of 0.021 in economics. For DiD-averaging the coefficient is \corrNeverCSA{} (\corrNeverCSAse{}). Never-treated units are clean controls and reduce TWFE's exposure to comparisons with already-treated units. Unlike in economics, later publication does not predict fewer reversals. The coefficients on the publication year are \corrYearCSA{} (\corrYearCSAse{}) and \corrYearBJS{} (\corrYearBJSse{}) per decade, and citing or reporting a modern estimator has no measurable association either. With \papersCSA{} and \papersBJS{} papers these regressions have little power, and we read them as descriptive.

\IfFileExists{tables/correlates.tex}{
\begin{table}[!htbp]\centering
\caption{Correlates of sign reversals}\label{tab:correlates}
\begin{threeparttable}
\begin{tabular}{lcc}
\toprule
 & DiD-averaging & Imputation \\
\midrule
Abstract/intro prominence & 0.140 (0.142) & 0.097 (0.146) \\
Always-treated share (per 10 pp) & -0.016 (0.041) & 0.073 (0.082) \\
Cites or reports a modern estimator & 0.065 (0.099) & -0.063 (0.095) \\
Event study & -0.085 (0.078) & 0.049 (0.080) \\
Log(1 + citations) & -0.033 (0.032) & 0.019 (0.031) \\
Marketing Science (vs MS) & -0.002 (0.089) & -0.079 (0.086) \\
Never-treated share (per 10 pp) & -0.012 (0.014) & -0.027 (0.013) \\
Preferred specification & -0.038 (0.102) & -0.056 (0.091) \\
Publication year / 10 & 0.280 (0.238) & -0.055 (0.243) \\
\bottomrule
\end{tabular}

\begin{tablenotes}\footnotesize
\item Notes: Each cell is the coefficient (paper-clustered standard error) from a separate regression of a sign-reversal indicator on the row variable, weighting each comparison by its inverse TWFE variance normalized to sum to one within paper, as in \citet{hull2026}. Always- and never-treated shares are measured in the regression sample and expressed per 10 percentage points. Unlike \citet{hull2026}, the publication-year and citation regressions include no journal or year fixed effects, ``cites or reports a modern estimator'' is one if the paper cites a modern-DiD methods paper or names one of the estimators in its full text, and journal indicators replace his journal-rank variables.
\end{tablenotes}
\end{threeparttable}
\end{table}
}{}

\subsection{Hierarchical model}
Following \citet{hull2026}, we fit a hierarchical model that separates the dispersion of TWFE and modern estimands from estimation noise, using the bootstrap covariance of each pair of estimates (Table~\ref{tab:hier}). The implied correlation between TWFE and modern estimands is \hCorrCSA{} for DiD-averaging and \hCorrBJS{} for imputation, against \corrCSA{} and \corrBJS{} for the estimates themselves. The prior probability of a sign reversal is \hFlipText{}, and the probabilities of a statistically significant TWFE estimate with the wrong sign are \hSigCSA{} and \hSigBJS{} percent, against 0.6 to 2.1 percent in economics. Single papers weigh heavily in these estimates, particularly for imputation, where a few precise estimates of opposite sign dominate the likelihood. The bootstrap standard errors of the estimand correlations, from resampling papers, are \hCorrCSAse{} for DiD-averaging and \hCorrBJSse{} for imputation. Dropping one paper at a time, the imputation estimand correlation ranges from \hCorrBJSlooMin{} to \hCorrBJSlooMax{} and the prior probability of a sign reversal from \hFlipBJSlooMin{} to \hFlipBJSlooMax{} percent. For DiD-averaging the ranges are \hCorrCSAlooMin{} to \hCorrCSAlooMax{} and \hFlipCSAlooMin{} to \hFlipCSAlooMax{} percent (Appendix Table~\ref{tab:loo}).

\IfFileExists{tables/hierarchical.tex}{
\begin{table}[!htbp]\centering
\caption{Hierarchical model estimates}\label{tab:hier}
\begin{threeparttable}
\begin{tabular}{lcc}
\toprule
 & DiD-averaging & Imputation \\
\midrule
SD of TWFE estimands & 0.134 & 0.131 \\
 & (0.021) & (0.035) \\
SD of modern-DiD estimands & 0.133 & 0.118 \\
 & (0.030) & (0.022) \\
Correlation of estimands & 0.618 & 0.608 \\
 & (0.119) & (0.287) \\
Prior probability of a sign reversal & 0.280 & 0.280 \\
 & (0.052) & (0.129) \\
Significant wrong-signed TWFE probability & 0.134 & 0.119 \\
 & (0.035) & (0.073) \\
\midrule
Observations & 283 & 306 \\
Papers & 28 & 29 \\
\bottomrule
\end{tabular}

\begin{tablenotes}\footnotesize
\item Notes: Gaussian precision-dependent hierarchical model as in \citet{hull2026}: estimand means and standard deviations depend on the demeaned log standard error of each estimator, with a common correlation parameter. Parameters are estimated by maximum likelihood with means profiled out by GLS. Probabilities average over the empirical distribution of standard errors. A sign reversal is $\theta_T\theta_M<0$; a significant wrong-signed TWFE estimate has $\hat\theta_T\theta_M<0$ and $|\hat\theta_T|>1.96 s_T$. Standard errors in parentheses come from 100 bootstrap resamples of papers.
\end{tablenotes}
\end{threeparttable}
\end{table}
}{\todo{Table: hierarchical model (generated by code/R/analysis.R once at least 30 comparisons are available).}}

\subsection{Comparison with economics}\label{sec:vshull}

\IfFileExists{tables/vs_hull.tex}{
\begin{table}[!htbp]\centering
\caption{Marketing and management versus economics}\label{tab:vshull}
\begin{threeparttable}
\footnotesize\setlength{\tabcolsep}{4pt}
\begin{tabular}{lcccc}
\toprule
 & \multicolumn{2}{c}{DiD-averaging} & \multicolumn{2}{c}{Imputation} \\
\cmidrule(lr){2-3}\cmidrule(lr){4-5}
 & This paper & Hull (2026) & This paper & Hull (2026) \\
\midrule
Correlation of estimates & 0.607 & 0.881 & 0.705 & 0.930 \\
Share of sign reversals & 0.251 & 0.145 & 0.180 & 0.108 \\
Share of significant sign reversals & 0.102 & 0.035 & 0.072 & 0.022 \\
Mean SE ratio (modern / TWFE) & 1.61 & 1.47 & 1.30 & 1.14 \\
Share of sign reversals, excl. pre-window averages & 0.264 & 0.128 & 0.167 & 0.083 \\
Correlation of estimands (hierarchical) & 0.618 & 0.913 & 0.608 & 0.976 \\
Prior probability of a sign reversal & 0.280 & 0.115 & 0.280 & 0.056 \\
Significant wrong-signed TWFE probability & 0.134 & 0.021 & 0.119 & 0.006 \\
\midrule
Comparisons & 283 & 1,069 & 306 & 1,116 \\
Papers & 28 & 82 & 29 & 89 \\
\bottomrule
\end{tabular}

\begin{tablenotes}\footnotesize
\item Notes: ``This paper'' columns reproduce Tables~\ref{tab:comparison} and \ref{tab:hier}. The Hull (2026) columns are taken from \citet{hull2026}. His estimates cover 1,069 DiD-averaging and 1,116 imputation comparisons from ten general-interest economics journals.
\end{tablenotes}
\end{threeparttable}
\end{table}
}{}

Correlations between TWFE and modern estimates are lower in our sample than in economics (Table~\ref{tab:vshull}). The gap is concentrated in unbalanced panels, as shown above, and the split by panel type that \citet{hull2026} provided gives a benchmark for each. In balanced panels and repeated cross-sections, our correlations of \corrCSAbal{} (DiD-averaging) and \corrBJSbal{} (imputation) are close to his 0.93 and 0.96. In unbalanced panels, ours are \corrCSAunbal{} and \corrBJSunbal{} against his 0.88 and 0.91, so imbalance alone does not account for the difference. In his sample the correlation is 0.06 lower in unbalanced panels for both estimators, with 95 percent confidence intervals of $[-0.14, 0.03]$ and $[-0.14, 0.02]$, and the ordering reverses when estimates are normalized by the TWFE standard error, as in his figures, rather than by the outcome standard deviation, as in all correlations reported here. In our sample the gap is larger and does not depend on the normalization. Three patterns are consistent with the lower agreement, although none amounts to a decomposition. First, negative weights carry more mass in our sample, and agreement falls where this mass is large, as in his sample. From the lowest to the highest tercile of the negative share of absolute weight, the correlation falls from \corrCSAnegLow{} to \corrCSAnegHigh{} for DiD-averaging and from \corrBJSnegLow{} to \corrBJSnegHigh{} for imputation. The two terciles hold \nNegLowCSA{} and \nNegHighCSA{} static comparisons for DiD-averaging and \nNegLowBJS{} and \nNegHighBJS{} for imputation. Second, agreement is also low in designs without never-treated units, in which TWFE identifies effects only from differences in adoption timing. Across the \nNoNeverCSA{} and \nNoNeverBJS{} such comparisons from \nNoNeverPapers{} papers, the correlations are \corrCSAnoNever{} and \corrBJSnoNever{}. Third, with \papersCSA{} and \papersBJS{} papers, against 82 and 89 in his sample, a few papers with many or very precise specifications weigh heavily on the correlation (Appendix Table~\ref{tab:loo}). Appendix Table~\ref{tab:disagreement} regresses sign reversals on these features jointly. For imputation, the absence of never-treated units predicts reversals, with a coefficient of \regNoNeverBJS{} (standard error \regNoNeverBJSse{}), and panel imbalance predicts them more weakly, with a coefficient of \regUnbalBJS{} (\regUnbalBJSse{}). For DiD-averaging, none of these features predicts reversals.

Adoption of the new estimators is recent in both journals. Among the \adoptMSN{} \emph{Management Science} papers with a staggered binary design, eligible or near-miss, none of the \adoptMSfirstN{} published in \adoptFirstPeriod{} reports a heterogeneity-robust estimator, against \adoptMSlastShare{} percent of the \adoptMSlastN{} published in \adoptLastPeriod{}. The first to do so appeared in \adoptMSfirstYear{}. In \emph{Marketing Science} none of the \adoptMKSCfirstN{} papers of \adoptFirstPeriod{} reports one, against \adoptMKSClastShare{} percent of the \adoptMKSClastN{} in \adoptLastPeriod{}, with the first in \adoptMKSCfirstYear{} (Appendix Table~\ref{tab:adoption}). The methods papers date from 2020 and 2021, so the first reports in these journals appear two to four years after them, which is consistent with the absence of a publication-year gradient in Table~\ref{tab:correlates}.

\subsection{Designs with time-varying controls}\label{sec:controls}

Time-varying controls, which the protocol excludes, are the most common reason a staggered design does not qualify. They are the first-listed violation for \nNearMissTVC{} of the \nNearMiss{} near-misses (Appendix~\ref{app:nearmiss}). To gauge whether the main results carry over to these designs, we re-estimate the papers whose staggered designs are reported with time-varying controls and whose packages contain usable data, mostly near-misses whose only violation is the controls. TWFE keeps the published controls, and the imputation estimator includes them in its first stage. The DiD-averaging estimator conditions on each control's first observed value in panels, balanced or not (outcome regression), and on period-demeaned controls in repeated cross-sections (the doubly robust estimator of \citealp{callaway2021}). They contribute \nCtrlRows{} specifications from \nCtrlPapers{} papers. After the leverage screen, \ctrlNBJS{} imputation comparisons from \ctrlPapersBJS{} papers and \ctrlNCSA{} DiD-averaging comparisons from \ctrlPapersCSA{} papers remain. The screen removes \ctrlCSAlevFail{} DiD-averaging comparisons from \ctrlCSAlevFailPapers{} papers whose TWFE estimates pass it: with the controls, the cell-level regressions put a large share of the weight on a single cluster, and the estimates have standard errors several times those of TWFE.

Table~\ref{tab:controls} reports the comparisons. Imputation estimates with controls track TWFE closely. The correlation is \ctrlCorrBJS{}, estimates change sign in \ctrlFlipBJS{} percent of comparisons, and in \ctrlSigBJS{} percent the difference is also significant. DiD-averaging estimates diverge more, with a correlation of \ctrlCorrCSA{} and sign reversals in \ctrlFlipCSA{} percent of comparisons, and their standard errors are \ctrlSeCSA{} times those of TWFE on average. Part of the gap reflects how the estimators condition on the controls. TWFE and imputation condition on their contemporaneous values, which can bias TWFE even with two periods when the controls respond to the treatment or their effects change over time \citep{caetano2022}, whereas DiD-averaging conditions on their pre-treatment values, as \citet{callaway2021} recommend. The gap therefore reflects the different conditioning as well as the change of estimator. Dropping specifications whose controls are plausible mediators of the treatment does not change these conclusions.

\IfFileExists{tables/controls.tex}{
\begin{table}[!htbp]\centering
\caption{TWFE versus modern DiD estimates in designs with time-varying controls}\label{tab:controls}
\begin{threeparttable}
\footnotesize\setlength{\tabcolsep}{3pt}
\begin{tabular}{llrrrrrr}
\toprule
Sample & Estimator & Comparisons & Papers & Corr. & Flips (\%) & Sig.\ flips (\%) & SE ratio \\
\midrule
All specifications & DiD-averaging & 90 & 10 & 0.80 & 28.9 & 14.4 & 2.46 \\
All specifications & Imputation & 100 & 11 & 0.99 & 3.0 & 1.0 & 1.04 \\
No mediator controls & DiD-averaging & 84 & 9 & 0.81 & 25.0 & 10.7 & 2.50 \\
No mediator controls & Imputation & 94 & 10 & 0.99 & 3.2 & 1.1 & 1.05 \\
\bottomrule
\end{tabular}

\begin{tablenotes}\footnotesize
\item Notes: Specifications reported with time-varying controls, mostly from near-miss papers whose only violation of the protocol is the controls (Appendix~\ref{app:nearmiss}). TWFE includes the published controls, and the imputation estimator includes them in the first stage. The DiD-averaging estimator conditions on each control's first observed value in panels, balanced or not, and on period-demeaned controls with the doubly robust estimator in repeated cross-sections. ``No mediator controls'' drops specifications whose controls are plausibly affected by the treatment. Same statistics and leverage screen as Table~\ref{tab:comparison}. The tests use the bootstrap covariance for imputation; for DiD-averaging with controls the bootstrap is not run and the covariance is zero (Section~\ref{sec:methods}).
\end{tablenotes}
\end{threeparttable}
\end{table}
}{}

\section{Conclusion}\label{sec:conclusion}
We re-estimate \nMainSpecs{} published staggered-adoption specifications from \nMainPapers{} papers in \emph{Management Science} and \emph{Marketing Science} with the DiD-averaging and imputation estimators, following the protocol of \citet{hull2026}, and compare estimators on the \nCSA{} and \nBJS{} comparisons from \papersCSA{} and \papersBJS{} papers that pass its leverage screen. We draw three conclusions. First, where staggered designs in business-school journals can be re-estimated on original data, TWFE and heterogeneity-robust estimators disagree more often than in economics. Correlations are lower, and sign reversals are more frequent, including statistically significant ones, at \sigflipCSA{} and \sigflipBJS{} percent. The gap is concentrated in unbalanced panels, in which units enter and exit, and in designs without never-treated units, and a few papers account for much of it. Second, the heterogeneity-robust estimators cost precision. Modern standard errors are on average \seratioCSA{} and \seratioBJS{} times those of TWFE, against 1.47 and 1.14 in \citet{hull2026}. The medians are \seratioCSAmed{} and \seratioBJSmed{}, so the larger means reflect a few specifications, mostly in unbalanced panels and in designs that impute untreated outcomes far beyond the last untreated observation. Third, the reanalysis itself is constrained by publication practice. Time-varying controls are the most common reason staggered designs fail the protocol, and \nEligSynth{} of the \nEligible{} papers with an eligible design come with synthetic or disguised data. Where designs with controls can be re-estimated, the imputation estimator with the same controls stays close to TWFE, while DiD-averaging, which conditions on the controls' pre-treatment values, diverges more. Both practices limit what readers and replicators can learn from published staggered designs.

Two limitations follow from the sample. With \papersCSA{} and \papersBJS{} papers from two journals behind the main comparisons, single papers move the summary statistics, as Appendix~\ref{app:loo} shows. The screening relied on language model agents working from article text and packages. The check of \nScreenedNonCand{} articles outside the candidate list added \nEligibleNonCand{} eligible designs, so the candidate search is not exhaustive. For authors of staggered designs in these journals, the results suggest reporting a heterogeneity-robust estimate alongside TWFE, in particular in unbalanced panels and in designs without never-treated units, where the two disagree most.

\clearpage
\bibliographystyle{aer}
\bibliography{refs}

\clearpage
\appendix
\section*{Appendix}
\section{Replication Protocol}\label{app:protocol}

The protocol adapts that of \citet{hull2026}. This appendix summarizes the eligibility rules, the estimation steps, and the rulings we made on cases the rules do not settle. The full protocol is part of our replication repository.

\subsection*{Eligibility}

We keep a paper if it reports at least one eligible specification in the main text (tables, figures, or numerical prose) and that specification can be replicated from the public code and data in its replication package. A specification is eligible if it satisfies four conditions.
\begin{enumerate}
\item The treatment is binary and switches on at unit-specific adoption dates. At least two distinct first-adoption dates occur among cohorts with both untreated and treated observations, and an untreated comparison unit exists in at least one treated period.
\item The specification is a linear regression with unit (or group) fixed effects and calendar-time fixed effects: a static TWFE regression or an unstacked event study with relative-time indicators.
\item The specification may include analysis weights and time-invariant characteristics interacted with calendar-time fixed effects (for example, region-by-year fixed effects), and the data may be a panel or repeated cross-sections assigned to stable units.
\item The specification includes none of the following: time-varying controls, lagged outcomes, unit-specific trends, additional treatments, treatment-intensity or continuous regressors, subgroup-by-treatment interactions, other nonredundant additive fixed effects, stacking, instrumental variables, nonlinear models (Poisson, logit), synthetic controls, or matching-based DiD.
\end{enumerate}
Reported DiD-averaging or imputation estimates of a design that meets these restrictions are also eligible, as are Sun--Abraham interacted event studies when they coincide with the DiD-averaging estimator. When the treatment switches off, we drop the post-adoption periods in which it equals zero.

The unit of analysis is every qualifying main-text specification, regardless of sign, size, significance, or the authors' interpretation. Specifications that differ in outcome, outcome transformation, sample, treatment definition, event window or binning, weights, fixed effects, or controls count separately, and a jointly estimated event-study path is one specification.

\subsection*{Estimation}

For each eligible specification, we proceed in eight steps.
\begin{enumerate}
\item Replicate the published TWFE estimate from the package and record any deviation.
\item Standardize the outcome to unit standard deviation in the regression sample.
\item Estimate TWFE with the original weights, fixed effects, and clustering.
\item Estimate the DiD-averaging estimator: cohort-by-period average treatment effects on the treated, with never-treated and not-yet-treated units as controls, aggregated with convex weights proportional to the number of treated observations. Event-study versions pool the cohort-by-period effects within the published bins and subtract the effect in the reference period. For papers that publish a Callaway--Sant'Anna estimate, also estimate it with the authors' comparison group, cells, aggregation and cohort weights, as a check (Appendix~\ref{app:pubcs}).
\item Estimate the imputation estimator: fit the fixed-effects model on untreated observations, impute untreated outcomes for treated cells, and average the differences. Pre-trend summaries follow \citet{hull2026}.
\item Estimate covariances between estimators with a joint weighted (Bayesian) bootstrap: 999 draws of unit-exponential weights at the cluster level, with TWFE and both modern estimators re-estimated on each draw. The unweighted draw must reproduce the main estimates, which we check for every specification.
\item Apply the leverage screen of \citet{hull2026}: write each estimate as a weighted sum of the outcomes, sum the squared weights within each cluster, and divide by their total to obtain each cluster's leverage share; the effective number of clusters is the inverse of the sum of squared shares. Keep a pair when both estimators have at least 10 effective clusters and a maximum share of at most 0.25.
\item Compute the ex post TWFE weights of \citet{borusyak2024aea}: the share of treated observations with a negative weight and the negative share of absolute weight.
\end{enumerate}

\subsection*{Rulings on judgment calls}

We resolved the following cases as of November 4, 2026.
\begin{itemize}
\item \emph{Matched samples.} A plain TWFE regression on a matched sample (for example, unweighted one-to-one nearest-neighbor matching) counts as a sample restriction and is eligible. Matching estimators, inverse-probability weighting, and synthetic controls are not.
\item \emph{Random effects.} Random-effects intercepts estimated by restricted maximum likelihood in place of fixed effects are not eligible as reported.
\item \emph{Unit trends.} Assignment-unit-specific linear trends in every reported column make the paper ineligible, following the unit-trend rule of \citet{hull2026}.
\item \emph{Time-varying controls.} A single time-varying control in every reported column makes the paper ineligible, and we record such papers as near-misses (Appendix~\ref{app:nearmiss}).
\item \emph{Extra treatment regressors.} Anticipation indicators, release indicators, and trends in time since adoption make a column ineligible, as do triple differences with an additional staggered regressor, such as a comparison of the same unit across platforms.
\item \emph{Additional fixed effects.} Additive fixed effects beyond unit and (group-by-)time effects, such as age fixed effects, pair-level post main effects, or event-time main effects for all units, make a column ineligible.
\item \emph{Anticipation bins.} A single anticipation bin next to the post indicator (for example, event months $-2$ and $-1$) is treated as a two-bin event study with the published bins. We do not construct a static version that the paper does not report.
\item \emph{Specifications not in the article.} Specifications that the published article does not report, such as columns that appear only in the package code, are excluded, as are published estimates that the package code does not reproduce. Reference values are the values printed in the published article, in its tables, text, or printed appendix. Estimates shown only in a figure have no reference value and are checked against the authors' code output. No value was read off a figure.
\item \emph{Treatments that switch off.} We keep the paper and drop post-adoption periods in which the treatment has switched off \citep{hull2026}.
\item \emph{Reported modern estimators.} We admit the DiD-averaging estimator of \citet{callaway2021}, the \citet{sun2021} estimator when it coincides with it, the imputation estimator of \citet{borusyak2024}, and Gardner's two-stage estimator. Extended TWFE (Wooldridge), synthetic DiD, matrix completion, and stacked DiD are not admitted, and a paper that reports only these estimators is a near-miss.
\item \emph{Main text only.} We do not count qualifying specifications that appear only in an appendix, but we list them in the screening record; Appendix~\ref{app:appendixspecs} repeats the comparison with the specifications printed in the articles' appendices added.
\item \emph{Working papers.} When screening relied on a working-paper version, we recheck eligibility and table references against the published article once it is available.
\item \emph{Data provenance.} Random subsamples and linear rescalings of the original data made by the authors count as original data and are flagged. Synthetic, disguised, pseudo, scrambled, or fuzzed data are excluded from the main analysis.
\item \emph{Unbalanced panels.} As in \citet{hull2026}, the DiD-averaging step uses cohort and comparison-group sample averages of whichever units are observed in each period: \texttt{att\_gt} from the \texttt{did} package with its unbalanced-panel option, which treats each cohort-by-period cell as a repeated cross-section, with the package's standard errors. For balanced panels this coincides with the panel estimator.
\item \emph{Repeated cross-sections.} Data with more than one observation per unit or group and period are treated as repeated cross-sections, following the notation of \citet{hull2026}, in which observations are indexed separately from the units or groups that carry the fixed effects. In two papers, a few unit-periods are duplicated by data construction: a merge repeats the rows of a few units in one, and overlapping event windows repeat a small share of unit-periods in the other. We collapse each duplicated unit-period to one row, with the mean outcome and a frequency weight, which leaves the TWFE and imputation point estimates unchanged, and estimate the data as a panel. Every other specification with duplicated unit-periods has them in at least 10 percent of unit-periods and is treated as repeated cross-sections.
\item \emph{High-cardinality characteristics.} Time-invariant characteristics interacted with time effects enter the DiD-averaging estimator as category indicators. A specification whose characteristic has more than 60 categories gets no DiD-averaging estimate and is flagged, because with all categories included the DiD-averaging cell-level regressions are not identified, and dropping the characteristic from DiD-averaging alone would compare estimators that adjust for different covariates (Section~\ref{sec:methods}). TWFE and imputation keep the authors' regression for these specifications.
\item \emph{Computational limits.} Where the problem is too large for the \texttt{did} package, we compute the DiD-averaging estimator without the package's influence-function matrix. Without covariates, we compute it directly from cohort and comparison-group means, with bootstrap standard errors. With covariates, we compute it cell by cell from the same doubly robust or regression cell estimates and accumulate the influence function of the reported summary as a single vector. Both implementations reproduce the package's estimates and standard errors wherever the package runs. No comparison is dropped for computational reasons.
\end{itemize}

\section{Results Without the Leverage Screen}\label{app:noscreen}

Following \citet{hull2026}, the main comparisons keep a pair of estimates only when both estimators have at least 10 effective clusters and a maximum cluster share of at most 25 percent. The screen removes every specification of \nNoPassPapers{} papers, whose treatments vary across a few states, adoption cohorts, or other clusters, and some specifications of other papers. Table~\ref{tab:noscreen} reports the comparisons without the screen: \nCSAnoscreen{} DiD-averaging comparisons from \papersCSAnoscreen{} papers and \nBJSnoscreen{} imputation comparisons from \papersBJSnoscreen{} papers. The correlations are \corrCSAnoscreen{} and \corrBJSnoscreen{} (\corrCSA{} and \corrBJS{} with the screen). Estimates change sign in \flipCSAnoscreen{} and \flipBJSnoscreen{} percent of comparisons, and in \sigflipCSAnoscreen{} and \sigflipBJSnoscreen{} percent the difference is also significant.

\IfFileExists{tables/noscreen.tex}{
\begin{table}[!htbp]\centering
\caption{TWFE versus modern DiD estimates with and without the leverage screen}\label{tab:noscreen}
\begin{threeparttable}
\begin{tabular}{lcccc}
\toprule
 & \multicolumn{2}{c}{DiD-averaging} & \multicolumn{2}{c}{Imputation} \\
\cmidrule(lr){2-3}\cmidrule(lr){4-5}
Leverage screen & Applied & Not applied & Applied & Not applied \\
\midrule
Correlation of estimates & 0.607 & 0.632 & 0.705 & 0.802 \\
Share of sign reversals & 0.251 & 0.246 & 0.180 & 0.161 \\
Share of significant sign reversals & 0.102 & 0.097 & 0.072 & 0.061 \\
Mean SE ratio (modern / TWFE) & 1.61 & 1.78 & 1.30 & 1.32 \\
\midrule
\multicolumn{5}{l}{\emph{Paper-weighted (each paper counts once)}} \\
Correlation of estimates & 0.621 & 0.756 & 0.350 & 0.558 \\
Share of sign reversals & 0.199 & 0.209 & 0.164 & 0.139 \\
Share of significant sign reversals & 0.112 & 0.108 & 0.115 & 0.090 \\
\midrule
Comparisons & 283 & 350 & 306 & 411 \\
Papers & 28 & 35 & 29 & 40 \\
\bottomrule
\end{tabular}

\begin{tablenotes}\footnotesize
\item Notes: Same statistics as Table~\ref{tab:comparison}. ``Applied'' is the main sample, in which both estimators have at least 10 effective clusters and a maximum cluster share of at most 25 percent. ``Not applied'' keeps every comparison with both estimates. Comparisons without a DiD-averaging estimate (a characteristic with more than 60 categories, or cohort-by-period cells not supported; Section~\ref{sec:methods}) are excluded from the DiD-averaging columns.
\end{tablenotes}
\end{threeparttable}
\end{table}
}{}

\section{Sensitivity to Individual Papers}\label{app:loo}

Because the comparisons come from \papersCSA{} and \papersBJS{} papers, single papers with many specifications or with very precise estimates can move the summary statistics. Table~\ref{tab:loo} reports each statistic for the full sample and its range when the comparisons of one paper at a time are dropped. Every paper is treated alike, and we do not exclude any paper from the main results.

\IfFileExists{tables/loo.tex}{
\begin{table}[!htbp]\centering
\caption{Summary statistics dropping one paper at a time}\label{tab:loo}
\begin{threeparttable}
\footnotesize\setlength{\tabcolsep}{4pt}
\begin{tabular}{lcccc}
\toprule
 & \multicolumn{2}{c}{DiD-averaging} & \multicolumn{2}{c}{Imputation} \\
\cmidrule(lr){2-3}\cmidrule(lr){4-5}
 & All & One dropped & All & One dropped \\
\midrule
Correlation of estimates & 0.607 & [0.480, 0.735] & 0.705 & [0.589, 0.846] \\
Share of sign reversals & 0.251 & [0.213, 0.267] & 0.180 & [0.155, 0.192] \\
Share of significant sign reversals & 0.102 & [0.080, 0.117] & 0.072 & [0.062, 0.086] \\
\midrule
\multicolumn{5}{l}{\emph{Paper-weighted (each paper counts once)}} \\
Correlation of estimates & 0.621 & [0.527, 0.667] & 0.350 & [0.173, 0.876] \\
Share of sign reversals & 0.199 & [0.169, 0.206] & 0.164 & [0.134, 0.170] \\
Share of significant sign reversals & 0.112 & [0.079, 0.116] & 0.115 & [0.083, 0.119] \\
\midrule
\multicolumn{5}{l}{\emph{Hierarchical model}} \\
Correlation of estimands & 0.618 & [0.545, 0.712] & 0.608 & [0.482, 0.921] \\
Prior probability of a sign reversal & 0.280 & [0.241, 0.308] & 0.280 & [0.126, 0.332] \\
Significant wrong-signed TWFE prob. & 0.134 & [0.109, 0.157] & 0.119 & [0.032, 0.154] \\
\midrule
Papers & 28 & & 29 & \\
\bottomrule
\end{tabular}

\begin{tablenotes}\footnotesize
\item Notes: ``All'' reproduces Tables~\ref{tab:comparison} and \ref{tab:hier}. ``One dropped'' reports the minimum and maximum of each statistic over samples that drop the comparisons of one contributing paper at a time. The hierarchical model is re-estimated on each sample.
\end{tablenotes}
\end{threeparttable}
\end{table}
}{}

\section{Design Features and Sign Reversals}\label{app:disagreement}

Table~\ref{tab:disagreement} regresses a sign-reversal indicator on design features jointly, with the weights of Table~\ref{tab:correlates}. The always-treated share and the negative share of absolute TWFE weight are defined for static specifications only and enter the static columns.

\IfFileExists{tables/disagreement.tex}{
\begin{table}[!htbp]\centering
\caption{Design features and sign reversals}\label{tab:disagreement}
\begin{threeparttable}
\footnotesize\setlength{\tabcolsep}{4pt}
\begin{tabular}{lcccc}
\toprule
 & \multicolumn{2}{c}{DiD-averaging} & \multicolumn{2}{c}{Imputation} \\
\cmidrule(lr){2-3}\cmidrule(lr){4-5}
 & All & Static & All & Static \\
\midrule
Unbalanced panel & -0.013 (0.159) & -0.047 (0.173) & 0.115 (0.060) & 0.042 (0.040) \\
No never-treated units & 0.107 (0.138) & 0.148 (0.160) & 0.453 (0.195) & 0.415 (0.153) \\
Log adoption cohorts & -0.047 (0.047) & -0.018 (0.061) & -0.020 (0.036) & -0.006 (0.044) \\
Event-study window & -0.062 (0.073) &  & 0.045 (0.091) &  \\
Always-treated share (per 10 pp) &  & 0.002 (0.060) &  & 0.030 (0.094) \\
Negative weight share (per 10 pp) &  & -0.027 (0.085) &  & 0.075 (0.064) \\
\midrule
Mean sign reversal (weighted) & 0.190 & 0.212 & 0.150 & 0.134 \\
Comparisons & 283 & 161 & 306 & 183 \\
Papers & 28 & 26 & 29 & 26 \\
\bottomrule
\end{tabular}

\begin{tablenotes}\footnotesize
\item Notes: Linear probability models of a sign reversal between TWFE and the modern estimate, estimated on the comparisons of Table~\ref{tab:comparison}. Each comparison is weighted by its inverse TWFE variance, normalized to sum to one within paper, as in \citet{hull2026}, and standard errors in parentheses are clustered by paper. ``All'' uses every comparison; ``Static'' uses static specifications only and adds the always-treated share and the negative share of absolute TWFE weight.
\end{tablenotes}
\end{threeparttable}
\end{table}
}{}

\section{Near-Miss Designs}\label{app:nearmiss}

We define a near-miss as a paper with a staggered binary adoption design and unit and time fixed effects in which every reported main-text column violates at least one restriction of Appendix~\ref{app:protocol}, so that no specification qualifies. Section~\ref{sec:controls} re-estimates the near-misses whose only violation is time-varying controls, with the controls in the imputation first stage and as baseline covariates in the DiD-averaging estimator. Table~\ref{tab:nearmiss} classifies near-misses by the restriction their screening record lists first.%
\IfFileExists{tables/near_miss_numbers.tex}{ Of the \nNearMiss{} near-misses (\shareNearMiss{} percent of the \nScreened{} screened papers), time-varying controls are the first-listed violation for \nNearMissTVC{}, and \nNearMissSole{} list a single violation. The keyword screen that assembles the near-miss list also returns \nNearMissExcluded{} papers with a single adoption date, a continuous treatment, or no panel DiD, and \nNearMissEligExcl{} papers that have a qualifying specification but no usable data. Neither the papers without a staggered binary design nor the eligible designs without usable data are near-misses, and Table~\ref{tab:nearmiss} reports both groups in memo rows. Two checks were made record by record. First, 11 records screened before the final coding rule, under which a staggered design whose every main-text column violates a rule is a near-miss rather than an eligible design, were re-read, and 9 became near-misses. Second, for 6 records the keyword classification and the screening agent disagreed on whether time-varying controls are the only violation, and each was settled by re-reading the record. The screening records were corrected accordingly.}{}

\IfFileExists{tables/near_miss_summary.tex}{
\begin{table}[!htbp]\centering
\caption{Near-miss papers by reason for exclusion}\label{tab:nearmiss}
\begin{threeparttable}
\small
\begin{tabular}{lccc}
\toprule
 & & No other & Percent of \\
 & Papers & violation listed & screened papers \\
\midrule
Time-varying controls & 90 & 50 & 6.8 \\
Extra treatment regressors or interactions & 31 & 7 & 2.3 \\
Additional fixed effects (age, event time) & 4 & 2 & 0.3 \\
Unit-specific trends & 4 & 1 & 0.3 \\
Stacked design & 25 & 4 & 1.9 \\
Matching, IPW or entropy-balancing weights & 14 & 2 & 1.1 \\
Nonlinear model (Poisson, logit, PPML) & 9 & 1 & 0.7 \\
ETWFE or other non-admitted estimator only & 6 & 2 & 0.5 \\
Other & 3 & 3 & 0.2 \\
\midrule
All near-miss papers & 186 & 72 & 14.0 \\
\midrule
\multicolumn{4}{l}{\emph{Memo: excluded keyword-screen records}} \\
No staggered binary design & 160 & -- & 12.0 \\
Eligible design, data not usable & 3 & -- & 0.2 \\
\midrule
Screened papers & 1,331 & & \\
\bottomrule
\end{tabular}

\begin{tablenotes}\footnotesize
\item Notes: Each near-miss is assigned the restriction its screening record lists first, using regular expressions on the free-text reason. ``No other violation listed'' counts records whose reason matches no other category. Near-misses include papers the screener flagged as near-misses and papers with a staggered binary design that a keyword screen for time-varying controls or extra treatment regressors adds. Percentages are relative to the \nScreened{} screened papers of Table~\ref{tab:funnel}.
\end{tablenotes}
\end{threeparttable}
\end{table}
}{}

\section{Adoption of Modern Estimators}\label{app:adoption}

Table~\ref{tab:adoption} reports, by publication year and journal, the share of screened papers with a staggered binary design (eligible designs and near-misses together) that report at least one heterogeneity-robust estimator.

\IfFileExists{tables/adoption.tex}{
\begin{table}[!htbp]\centering
\caption{Papers with a staggered design that report a modern estimator, by year}\label{tab:adoption}
\begin{threeparttable}
\begin{tabular}{lcccc}
\toprule
 & \multicolumn{2}{c}{\emph{Management Science}} & \multicolumn{2}{c}{\emph{Marketing Science}} \\
\cmidrule(lr){2-3}\cmidrule(lr){4-5}
Publication year & Papers & Share modern & Papers & Share modern \\
\midrule
2013--2018 & 0 & -- & 5 & 0.000 \\
2019--2022 & 29 & 0.000 & 14 & 0.000 \\
2023--2024 & 53 & 0.113 & 6 & 0.500 \\
2025--2026 & 89 & 0.326 & 15 & 0.267 \\
\midrule
All years & 171 & 0.205 & 40 & 0.175 \\
\bottomrule
\end{tabular}

\begin{tablenotes}\footnotesize
\item Notes: Screened papers with an eligible staggered design or a near-miss design (Appendix~\ref{app:nearmiss}), by journal and publication year. ``Share modern'' is the share whose screening record lists a Callaway--Sant'Anna, Sun--Abraham, Borusyak--Jaravel--Spiess, Gardner, or de Chaisemartin--D'Haultf\oe{}uille estimate. Papers screened from the article text alone (Table~\ref{tab:funnel}) are not included, because their screening records carry no estimator field. Years are grouped so that each cell holds several papers.
\end{tablenotes}
\end{threeparttable}
\end{table}
}{}

\section{Papers That Publish a Callaway--Sant'Anna Estimate}\label{app:pubcs}

The main analysis applies the DiD-averaging settings of \citet{hull2026} to every specification. For the papers that publish a Callaway--Sant'Anna estimate, Table~\ref{tab:pubcs} compares our estimates with the authors' settings against the published estimates and against our main estimates. Table~\ref{tab:authcs} repeats the main DiD-averaging comparison with the authors' settings in place of Hull's for these papers.

\IfFileExists{tables/published_cs.tex}{\begin{table}[htbp]\centering
\caption{DiD-averaging estimates of papers that publish a Callaway--Sant'Anna estimate}\label{tab:pubcs}
\footnotesize
\setlength{\tabcolsep}{4pt}
\begin{tabular}{lcccccccc}
\toprule
 & & & \multicolumn{3}{c}{Authors' settings vs.\ published} & \multicolumn{3}{c}{Hull's vs.\ authors' settings} \\
\cmidrule(lr){4-6}\cmidrule(lr){7-9}
 & & & & & Max & Median & Max & Sign \\
Authors' software & Papers & Specs. & Printed & Matched & gap/SE & gap/SE & gap/SE & changes \\
\midrule
R \texttt{did} or \texttt{staggered} & 5 & 70 & 44 & 42 & 0.07 & 0.00 & 4.2 & 0 \\
Stata \texttt{csdid} & 3 & 10 & 10 & 10 & 0.04 & 0.17 & 5.5 & 3 \\
\midrule
All & 8 & 80 & 54 & 52 & 0.07 & 0.01 & 5.5 & 3 \\
\bottomrule
\end{tabular}
\par\smallskip
\begin{minipage}{0.95\textwidth}
{\footnotesize\noindent Notes: Papers that publish a Callaway--Sant'Anna estimate, grouped by the software of the published estimate; specifications count static estimates and event-study post-window averages. ``Printed'' counts specifications with a printed estimate (some event studies are shown only in figures); ``Matched'' counts the specifications for which our estimate under the authors' implementation choices matches the printed estimate to the last printed decimal. Gaps to the published estimate are in units of the published standard error. For 6 specifications of one paper, the protocol drops post-adoption rows in which the treatment switches off; their reference is the authors' estimator re-run on that sample, since the printed estimate, which their code reproduces on the full sample, is not the target. The last three columns compare these estimates with our main estimates, which use the settings of \citet{hull2026}, and report the absolute difference in units of the standard error of the main estimate.\par}
\end{minipage}
\end{table}
}{}

\IfFileExists{tables/authcs.tex}{
\begin{table}[!htbp]\centering
\caption{DiD-averaging with Hull's and with the authors' settings}\label{tab:authcs}
\begin{threeparttable}
\begin{tabular}{lcccc}
\toprule
 & \multicolumn{2}{c}{All specifications} & \multicolumn{2}{c}{Papers publishing a CS estimate} \\
\cmidrule(lr){2-3}\cmidrule(lr){4-5}
CS settings & Hull's & Authors' & Hull's & Authors' \\
\midrule
Correlation of estimates & 0.607 & 0.604 & 0.478 & 0.469 \\
Share of sign reversals & 0.251 & 0.240 & 0.247 & 0.215 \\
Share of significant sign reversals & 0.102 & 0.092 & 0.161 & 0.129 \\
Mean SE ratio (modern / TWFE) & 1.61 & 1.60 & 1.62 & 1.59 \\
\midrule
\multicolumn{5}{l}{\emph{Paper-weighted (each paper counts once)}} \\
Correlation of estimates & 0.621 & 0.620 & 0.784 & 0.776 \\
Share of sign reversals & 0.199 & 0.163 & 0.276 & 0.151 \\
Share of significant sign reversals & 0.112 & 0.076 & 0.192 & 0.067 \\
\midrule
Comparisons & 283 & 283 & 93 & 93 \\
Papers & 28 & 28 & 8 & 8 \\
\bottomrule
\end{tabular}

\begin{tablenotes}\footnotesize
\item Notes: Same statistics as Table~\ref{tab:comparison}, for the DiD-averaging estimator. The columns labeled Hull's are the main analysis, which applies the settings of \citet{hull2026} to every specification: not-yet-treated and never-treated comparison units, cohort averages of the units observed in each period, and weights proportional to the number of treated observations. The columns labeled Authors' use instead the authors' comparison group, cell construction, aggregation and cohort weights for the papers that publish a Callaway--Sant'Anna estimate. The last two columns restrict the sample to those papers.
\end{tablenotes}
\end{threeparttable}
\end{table}
}{}

\clearpage
\section{Specifications Printed in Appendices}\label{app:appendixspecs}

The protocol, like \citet{hull2026}, keeps only specifications reported in the main text. INFORMS journals print some appendices inside the article, and a few of those contain further staggered regressions: robustness variants of main-text specifications on matched samples, alternative windows or subsamples, and in one case the only staggered regressions of the paper. We screened the printed appendices of every paper in the replication sample, replicated the eligible columns under the same rules, and here repeat the comparison on the enlarged sample. Online supplements are not covered. \IfFileExists{tables_appx/results_numbers.tex}{The printed appendices add \nAddedSpecsAppx{} specifications from \nAddedPapersAppx{} papers, one of them new to the sample, and most of them from two papers. Table~\ref{tab:comparisonappx} reports the comparison on \nCSAAppx{} DiD-averaging and \nBJSAppx{} imputation comparisons from \papersCSAAppx{} and \papersBJSAppx{} papers. The correlations are \corrCSAAppx{} and \corrBJSAppx{}, against \corrCSA{} and \corrBJS{} in the main sample, and the reversal rates \flipCSAAppx{} and \flipBJSAppx{} percent, against \flipCSA{} and \flipBJS{}. Weighting papers equally, which limits the weight of the two papers that supply most of the added specifications, the reversal rates are \flipCSApwAppx{} and \flipBJSpwAppx{} percent.}{}

\IfFileExists{tables_appx/comparison.tex}{
\begin{table}[!htbp]\centering
\caption{TWFE versus modern DiD estimates, main sample plus specifications printed in appendices}\label{tab:comparisonappx}
\begin{threeparttable}
\begin{tabular}{lcc}
\toprule
 & DiD-averaging & Imputation \\
\midrule
Correlation of estimates & 0.562 & 0.725 \\
Share of sign reversals & 0.251 & 0.192 \\
Share of significant sign reversals & 0.102 & 0.066 \\
Mean SE ratio (modern / TWFE) & 1.61 & 1.27 \\
\midrule
\multicolumn{3}{l}{\emph{Paper-weighted (each paper counts once)}} \\
Correlation of estimates & 0.630 & 0.379 \\
Share of sign reversals & 0.197 & 0.168 \\
Share of significant sign reversals & 0.110 & 0.109 \\
\midrule
\multicolumn{3}{l}{\emph{Excluding pre-window averages of event studies}} \\
Correlation of estimates & 0.621 & 0.724 \\
Share of sign reversals & 0.262 & 0.176 \\
Share of significant sign reversals & 0.096 & 0.055 \\
Comparisons & 282 & 307 \\
\midrule
Comparisons & 342 & 365 \\
Papers & 29 & 30 \\
\bottomrule
\end{tabular}

\begin{tablenotes}\footnotesize
\item Notes: Same statistics, leverage screen, and significance test as Table~\ref{tab:comparison}, on the main sample plus the eligible specifications printed in the articles' appendices (Appendix~\ref{app:appendixspecs}).
\end{tablenotes}
\end{threeparttable}
\end{table}
}{}

\clearpage
\section{Papers in the Replication Sample}\label{app:papers}

Table~\ref{tab:paperlist} lists every paper with at least one re-estimated specification, with the number of specifications and whether the paper itself reports a modern DiD estimator.

\IfFileExists{tables/paper_list.tex}{\begingroup\scriptsize\setstretch{1}\renewcommand{\arraystretch}{0.95}
\setlength{\LTcapwidth}{\textwidth}
\begin{longtable}{rlcccc}
\caption{Papers in the replication sample}\label{tab:paperlist}\\
\toprule
 & & & & & Reports \\
 & Paper (DOI suffix) & Journal & Year & Specifications & modern DiD \\
\midrule
\endfirsthead
\multicolumn{6}{l}{\emph{Table \thetable, continued}} \\
\toprule
 & & & & & Reports \\
 & Paper (DOI suffix) & Journal & Year & Specifications & modern DiD \\
\midrule
\endhead
\midrule
\multicolumn{6}{r}{\emph{Continued on next page}} \\
\endfoot
\bottomrule
\endlastfoot
1 & \texttt{mksc.2017.1051} & MKSC & 2018 & 19 & No \\
2 & \texttt{mksc.2022.1356} & MKSC & 2022 & 9 & No \\
3 & \texttt{mksc.2022.1409} & MKSC & 2023 & 11 & Yes \\
4 & \texttt{mksc.2023.1437} & MKSC & 2023 & 20 & Yes \\
5 & \texttt{mnsc.2020.3626} & MS & 2021 & 9 & No \\
6 & \texttt{mnsc.2020.3707} & MS & 2021 & 1 & No \\
7 & \texttt{mnsc.2021.4043} & MS & 2022 & 2 & No \\
8 & \texttt{mnsc.2021.4057} & MS & 2022 & 6 & No \\
9 & \texttt{mnsc.2021.4092} & MS & 2022 & 1 & No \\
10 & \texttt{mnsc.2021.4103} & MS & 2022 & 5 & No \\
11 & \texttt{mnsc.2021.4209} & MS & 2022 & 1 & No \\
12 & \texttt{mnsc.2022.4326} & MS & 2022 & 11 & No \\
13 & \texttt{mnsc.2022.4394} & MS & 2023 & 64 & No \\
14 & \texttt{mnsc.2023.4716} & MS & 2023 & 3 & No \\
15 & \texttt{mnsc.2023.4884} & MS & 2023 & 10 & No \\
16 & \texttt{mnsc.2021.01740} & MS & 2024 & 2 & No \\
17 & \texttt{mnsc.2021.02026} & MS & 2024 & 11 & Yes \\
18 & \texttt{mnsc.2022.01820} & MS & 2024 & 18 & No \\
19 & \texttt{mnsc.2022.4660} & MS & 2024 & 2 & Yes \\
20 & \texttt{mnsc.2023.00900} & MS & 2024 & 1 & No \\
21 & \texttt{mnsc.2023.4800} & MS & 2024 & 2 & No \\
22 & \texttt{mnsc.2023.4816} & MS & 2024 & 8 & No \\
23 & \texttt{mnsc.2023.4968} & MS & 2024 & 8 & No \\
24 & \texttt{mnsc.2020.01306} & MS & 2025 & 7 & No \\
25 & \texttt{mnsc.2020.01838} & MS & 2025 & 30 & No \\
26 & \texttt{mnsc.2021.03484} & MS & 2025 & 3 & No \\
27 & \texttt{mnsc.2022.01333} & MS & 2025 & 4 & Yes \\
28 & \texttt{mnsc.2023.01357} & MS & 2025 & 2 & No \\
29 & \texttt{mnsc.2022.02310} & MS & 2026 & 4 & Yes \\
30 & \texttt{mnsc.2022.03186} & MS & 2026 & 12 & Yes \\
31 & \texttt{mnsc.2022.03268} & MS & 2026 & 1 & No \\
32 & \texttt{mnsc.2023.01390} & MS & 2026 & 4 & Yes \\
33 & \texttt{mnsc.2023.01423} & MS & 2026 & 10 & No \\
34 & \texttt{mnsc.2023.01572} & MS & 2026 & 14 & Yes \\
35 & \texttt{mnsc.2023.01987} & MS & 2026 & 1 & No \\
36 & \texttt{mnsc.2023.04090} & MS & 2026 & 4 & No \\
37 & \texttt{mnsc.2024.05423} & MS & 2026 & 2 & No \\
38 & \texttt{mnsc.2024.08465} & MS & 2026 & 1 & Yes \\
39 & \texttt{mnsc.2025.02515} & MS & 2026 & 30 & Yes \\
40 & \texttt{mnsc.2025.03238} & MS & 2026 & 2 & Yes \\
\end{longtable}
\vspace{-0.5\baselineskip}
{\footnotesize\noindent Notes: One row per paper with at least one re-estimated specification on original data, including random subsamples and linear rescalings of the original data made by the authors; papers whose packages contain synthetic, disguised, pseudo or simulated data are not re-estimated. Specifications count static estimates and event-study post-window averages (pre-window averages excluded), before the leverage screen. ``Reports modern DiD'' is based on the estimators listed in the screening record (Callaway--Sant'Anna, Sun--Abraham, Borusyak--Jaravel--Spiess, Gardner, or de Chaisemartin--D'Haultf\oe{}uille). MS = \emph{Management Science}; MKSC = \emph{Marketing Science}.\par}
\endgroup
}{}

\end{document}